\documentstyle[a4,cite,12pt,aps,epsfig,subeqn]{revtex}
\def\baselinestretch{1.2}
\makeatletter
\def\dgr{\dagger}
\def\nnb{\nonumber}

\def\be{\begin{equation}}
\def\ee{\end{equation}}
\def\mn{\mu\nu}
\def\don{d_1}
\def\dtw{d_2}
\def\dth{d_3}
\def\de{\delta}
\def\rx{{\stackrel{\rightharpoonup}{X}}}
\def\lx{{\stackrel{\leftharpoonup}{X}}}
\def\HH{{\overline H}}
\def\WW{{\overline W}}
\def\bare{{2 \over {\bar {\epsilon}}}}
\newcommand{\bea}{\begin{eqnarray}}
\newcommand{\eea}{\end{eqnarray}}

\def\jsep{\nnb \\ &&}
\def\ssep{\right. \nnb\\ && \left.}
\renewcommand{\thesection}{\arabic{section}}
\renewcommand{\theequation}{\thesection.\arabic{equation}}
\newcounter{saveeqn}

\makeatletter
\@addtoreset{equation}{section}

\makeatother

\begin{document}
\title{ One-loop anomalous couplings 
and  matching conditions  in the nonlinear realization of the SU(2)
Higgs Model } 

\author{
	{\bf Sukanta Dutta}\footnote{Email 
Address: Sukanta.Dutta@kek.jp \\
  (On leave from SGTB Khalsa College, University of
Delhi. Delhi-110007. India )},
	{\bf Kaoru Hagiwara }
           and  {\bf Qi-Shu Yan}\footnote{Email 
Address: yanqs@post.kek.jp}\\
Theory Group, KEK,  Tsukuba, 	 305-0801, Japan
}
\bigskip

\address{\hfill{}}

\maketitle
\begin{flushright}
\vskip -5.5 cm
{\bf KEK-TH-962 }
\vskip 5.5 cm
\end{flushright}
\begin{abstract}
We study the renormalization of the nonlinear realization of  the SU(2) Higgs
model in  the modified
minimal subtraction renormalization scheme. We propose that the effective
field method with truncated operator series
is trustworthy even when the Higgs boson is relatively light.
 Using the technique of background field  method in the coordinate
 space, we derive the matching conditions at
 tree and one loop order, both in the regions where the Higgs boson is 
 heavy and where it 
 is light. 
We obtain the complete one-loop anomalous
couplings up to ${\Large O}\!\! \left( p^4\right)$ in   the SU(2) Higgs
model.  We observe that the contribution 
of the gauge bosons and the Goldstone bosons  to the anomalous
couplings at 
the one-loop level  is significant.
  By establishing the correspondence between our coordinate space
  calculation 
and the momentum space calculation
that exists in the literature, we find agreement in all the matching 
conditions in the heavy Higgs boson limit.
\end{abstract}

\section{Introduction}
In this paper, we
study the renormalization of the $SU(2)$ 
Higgs model in the non-linear realization, 
and demonstrate how  the matching 
condition up to the one-loop level in the $SU(2)$ 
chiral Lagrangian is made, as prescribed in \cite{Georgi}.
Such a toy model has been
studied in the reference \cite{Dittmaier} 
to investigate the decoupling limit
of a heavy Higgs boson using the path integral method. 

The chiral Lagrangian method is 
one of the important derivatives 
of the effective field theory method, and
has been used for the description 
of both the hadronic \cite{gasser} 
and electroweak physics \cite{appliq}.
Considering the fact that the chiral 
Lagrangian for electroweak physics
contains more than ten extra operators, 
before examining the  chiral electroweak $SU(2) \times U(1)$ 
model, here we will  study the simplest
and non-trivial toy model: the $SU(2)$ Higgs 
model and its chiral effective Lagrangian
with spontaneous symmetry  breaking. 

We will use the modified minimal subtraction 
scheme (${\overline {MS}}$) in our calculation. 
It is a more  convenient  
renormalization scheme,  since the divergences can be automatically subtracted and
expressed in a  simple form.
With the right matching procedure in the ${\overline {MS}}$,
the infrared divergences of the 
effective theory are just  the same as those of 
the full theory. And
that's why it is regarded as one 
of the pillars of the effective field
theory method \cite{Georgi,Pich}.

According to the procedure specified 
for the effective field theory \cite{Georgi}, the matching conditions
are obtained order by 
order at the matching scale.  To the n-loop  order they can be expressed as
\bea
\int \delta {\cal L}^{n}(\phi) = S^n_{Full}(\phi,\Phi[\phi]) - S^n_{eff}(\phi)\,,
\label{mtce}
\eea
where $\phi$ corresponds to the degrees of freedom at low energy region,
and $\Phi$ corresponds to the heavy degrees of freedom, which at the
matching scale is expressed in terms of $\phi$ through its equation of motion (EOM)
. The terms  $S^n_{Full}(\phi,\Phi[\phi])$ and $S^n_{eff}(\phi)$ are
the effective actions of the full and the effective theories, respectively, up to 
$n$-loop order. The term $\delta {\cal L}^{n}(\phi)$ 
accounts for the n-loop matching conditions,
which determines the anomalous 
couplings at $n$-loop order. 

It is  known in  the literature, that there are two different (but
intrinsically the same) methods 
to construct the low energy effective
Lagrangian from a full renormalizable
theory. The first method is the standard
diagrammatic one, i.e. to compute the one particle 
irreducible Feynman diagrams ,
and then to match them order by order with those 
computed in the effective theory. 
The authors of \cite{boson} have used this method to extract
the non-decoupling terms for a 
heavy Higgs boson in the $SU(2) \times U(1)$ case.
And the authors of \cite{fermion} used 
it to consider one of the extensions of
the standard model with a heavy 
scalar singlet coupled to the leptonic doublet.

The second method is the path integral, in
which the effects of the heavy degrees of freedom 
can also be extracted order by order.
For example in the one-loop case, the heavy degrees of freedom
can be integrated out formally by the Gaussian
integration and we can  
compute the renormalization, purely with the algebraic 
manipulation and without any use of the
technique of Feynman diagrams. 
However, the second method is equivalent to
the first one, since it only combines 
the Feynman diagrams in a specific way (to guarantee
the gauge invariance in the gauge 
theories). 
We can still utilize the Feynman diagram method 
to guide our calculations. The authors of \cite{Dittmaier,Dittmaiersu2u1} 
have explored this method by deriving the non-decoupling
effects in the decoupling limit of a 
heavy Higgs boson in both the $SU(2)$ and 
$SU(2) \times U(1)$ cases. By using this method,
in the reference \cite{yandu},
the renormalization group equations (RGE) of the effective
chiral SU(2) Lagrangian are obtained, which will be used for
comparison in this article.

We would like to make some comments on 
these works: 
\begin{enumerate}
\item All previous studies on the chiral Lagrangian theories for the bosonic
  sector  derived the non-decoupling effect of the Higgs boson, in the
decoupling limit. In this limit one can evade  the evaluation
prescription for the matching conditions given by the Eq. (1), by
ignoring  the term  ${\cal L}^{1-loop}_{eff}$ which is suppressed by Higgs boson's mass.
However these studies are not applicable to the light Higgs boson
scenarios.  The authors of reference \cite{yandu} found that
when Higgs boson is relatively light, the predictions of the RGE
method deviate from the prediction of the decoupling limit significantly.

\item Although the references \cite{Dittmaier, Dittmaiersu2u1} worked
in the non-linear realization, they have not extracted all divergences
of the one loop effective Lagrangian. Therefore, the demonstration of the renormalizability
of these models in the non-linear realization is still missing.

\item The parameterization in the reference \cite{Dittmaier} 
generates 
D'Alambert operator for the Goldstone bosons given as
\bea
\Box^{ab}&=&\frac{(v + {\overline H})^2}{v^2} 
\Bigg\{ [- (d \cdot d)^{ab} 
+ m_W^2 \de^{ab} ] + \sigma^{ab}_{\xi} \Bigg\}\,,
\label{dao}
\eea 
which leads to an ill-defined propagator (where $m_W$ is the gauge boson
 mass and the definitions for
 $(d\cdot d)^{ab}$ and $ \sigma^{ab}_{\xi}$ are given later in
  section \ref{sec:su2hrenorm} and Eq.  (\ref{sigmaG}), respectively.
This ill-defined propagator introduces the quartic and the quadratic divergences 
and makes it difficult to verify the renormalizability of these models.
\end{enumerate}
Therefore, it is worthwhile  to study 
the renormalization of the non-linear 
$SU(2)$ Higgs model in the ${\overline {MS}}$ 
renormalization scheme,
and to derive the matching conditions up to one-loop 
level which is valid for a large range of Higgs boson's mass. 
We also examine the renormalization
scale dependence of the one-loop anomalous couplings. 

Just for the sake of convenience, we 
conduct all our calculation in the
Euclidean coordinate
space-time.  As the matter of fact, not only for the
divergent terms as mentioned  in \cite{Lee}, but also for
the finite terms, it is less cumbersome to work in the coordinate
space-time.

This paper is organized as follows.   Section \ref{sec:su2h},
 briefly introduces the Lagrangian ${\cal L}_{SU(2)}$ of 
the renormalizable $SU(2)$ Higgs model (full theory),
and concentrate on its nonlinear form. 
Section \ref{sec:effchiral}, introduces  the nonlinear effective $SU(2)$ Lagrangian
$L^{eff}$ up to $O(p^4)$.
In  section \ref{sec:su2hrenorm}, we perform the renormalization
of the ${\cal L}_{SU(2)}$ in the background field method (BFM). 
In  section \ref{sec:com1loop}, the complete one-loop anomalous couplings
in the SU(2) Higgs model are 
calculated. Section \ref{sec:treematch} presents the tree level and the
one-loop level matching conditions.
In  section \ref{sec:compare}, we list the 
the one-loop RGE calculation in the effective Lagrangian.
In  section \ref{sec:num}, numerical analysis is given to compare
the  effective couplings obtained by RGE method, those calculated in the  decoupling
limit \cite{Dittmaier} and those following from  the exact
one-loop calculation of the renormalizable SU(2) Higgs model.
We summarize this article with  discussions and conclusions. We
provide the basics of the short distance approximation in the appendix
\ref{app:heatkernel}.
In appendix \ref{app:div} and \ref{app:rge} we provide the necessary
mathematical tool to extract the divergent and the finite parts from
the effective generating functionals of SU(2) Higgs model and the effective SU(2) chiral
Lagrangian respectively. Appendix \ref{app:B0} addresses the definition
of the scalar loop integral in coordinate space formalism.

\section{The renormalizable $SU(2)$ Higgs model}
\label{sec:su2h}
The partition functional of the renormalizable
non-Abelian $SU(2)$ Higgs model 
\cite{higgs} (This does not include
the gauge fixing and  ghost terms. They will be added when necessary for the 
quantization of quantum fields ) can be expressed as
\bea
{\cal Z}=\int {\cal D}A_{\mu}^{a} {\cal D}\phi {\cal D}\phi^{\dgr}
\exp\left ({{\cal S}[A,\phi,\phi^{\dgr}]}\right )\,,
\eea
where the action ${\cal S}$ is 
determined by the following Lagrangian
density
\bea
\label{su2l}
{ \cal L} &=& -{1\over 4 g^2} W_{\mn}^a W^{a \mn}
	 - (D\phi)^{\dagger}\cdot(D \phi)
	+\mu^2 \phi^{\dagger} \phi - 
	{\lambda \over 4} (\phi^{\dagger} \phi)^2\,,
\label{linear}
\eea
and the definition of quantities in this Lagrangian is given below
\bea
W_{\mn}^a&=&\partial_{\mu} W_{\nu}^a - 
\partial_{\nu} W_{\mu}^a + f^{abc} W_{\mu}^b W_{\nu}^c\,,\\
D_{\mu} \phi&=&\partial_{\mu} \phi - i W_{\mu}^a T^a \phi\,,\\
\phi^{\dgr}&=&(\phi_1^*,\phi_2^*)\,,
\eea
where $T^a$ is the generator of the Lie algebra of
$SU(2)$ gauge group.

The spontaneous symmetry breaking is induced by
the positive mass square $\mu^2$ in the Higgs potential.
The vacuum expectation value of Higgs
field is given as $|\langle \phi \rangle| = v/{\sqrt 2}$.
And by eating the corresponding
Goldstone boson, the gauge bosons $W$ obtain their masses.

The non-linear realization of 
the Lagrangian given in Eq. (\ref{su2l})
is made by changing the variable $\phi$
\bea
\phi={1\over \sqrt{2}} (v + H) U\,,\,
U=\exp\left ( i { \zeta^a T^a }\right )\,,\,
v=2 \sqrt{\mu^2 \over \lambda}\,,
\eea
where the matrix field $U$ is the Goldstone boson field
as prescribed by the Goldstone theorem, 
$\zeta^a$ is the dimensionless
phase angle, and the $H$ is a
massive scalar field. Then the Lagrangian in Eq. (\ref{su2l}) is
rewritten as
\bea
{\cal L'} &=& -{1\over 4 g^2} W_{\mn}^a W^{a \mn}
	 - {(v+H)^2\over 4} tr[(D U)^{\dagger} \cdot (D U)] \nnb\\    
&&	- {1\over 2} \partial H \cdot \partial H
	+{1\over 2} \mu^2 (v+H)^2 - {\lambda \over 16} (v+H)^4\,,
\label{nonlinear}
\eea
where $tr$ is sum over the SU(2) group indices. 
And the change of variables induces a 
determinant factor in the functional integral ${\cal Z}$
\bea
{\cal Z} = \int {\cal D} W_{\mu}^a {\cal D} H {\cal D} 
\zeta^b \exp\left ({\cal S}'[W,H,\xi] \right ) 
\det\left \{ \left  (v + H\right  ) \delta(x-y) \right \}\,.
\eea
The determinant can be written in the exponential form, 
and correspondingly the Lagrangian density is modified to
\bea
{\cal L} \rightarrow {\cal L'} 
+ \delta(0) ln \left \{ v + H \right \}\,.
\eea
There exists an arbitrariness to express 
the phase angle $\zeta^a$ into a field. However, this
arbitarariness should not change the physics \cite{Haag}.
According to our experience in the 
hadronic chiral Lagrangian, the relation between
the phase angle $\zeta^a$ and the 
Goldstone field $\xi^a$ can be defined  as
\bea
\zeta^a &=& {{\bf 2 } \xi^a \over v}\,.
\eea
Then the partition function $Z$ can 
be expressed by quantum fields as
\bea
{\cal Z} = \int {\cal D} W_{\mu}^a {\cal D} H {\cal D} \xi^b 
\exp\left ({\cal S}'[W,H,\xi] \right ) 
\det\left \{ \left  (1 + {H \over v} \right  ) \delta(x-y) \right \}\,.
\eea
Now the Lagrangian density is modified to the following form
\bea
\label{hml}
{\cal L} \rightarrow {\cal L'} 
+ \delta(0) ln \left \{ 1 + {H \over v} \right \}\,.
\label{uglag}
\eea 
This determinant which
contains quartic divergences is
indispensable and crucial to cancel exactly the quartic
divergences brought into by the longitudinal part
of gauge boson, and is important in verifying the
renormalizability of the Higgs model in the U-gauge \cite{wood, ugauge}.

\section{The non-renormalizable effective 
chiral SU(2) theory up to $O(p^4)$}
\label{sec:effchiral}
Any effective Lagrangian 
is valid  within its infrared cutoff $\Lambda_{IR}$
and its ultraviolet cutoff $\Lambda_{UV}$.
In the nonlinear chiral effective $SU(2)$ Lagrangian $L^{eff}$,
the effective dynamic degrees of freedom at low energy region includes
only the Goldstone and the gauge bosons.  Masses of the quantum Goldstone bosons 
are gauge dependent, and can be vanishing or
infinitely heavy. But the masses of 
the gauge bosons are fixed by the experiments.
So the infrared cutoff $\Lambda_{IR}$ for this effective Lagrangian should
be larger than the masses of the gauge 
bosons at the low energy region in the theory. However, this fact invalidates the naive
momentum power counting rule in 
the hadronic chiral Lagrangian, where only massless
Goldstone bosons are included in the theory and the momentum
$p$ is assumed to be much smaller 
than the vacuum expectation value $v$.
For the ultraviolet cutoff $\Lambda_{UV}$
we know that it should be lower than the new resonances 
otherwise the new degrees of freedom would 
break the validity of  the effective theory.

In principle the Lagrangian $L^{eff}$,  which includes all permitted
operators composed by these light degrees of freedom
and respects the assumed Lorentz and gauge symmetries, is
still renormalizable \cite{wein}. But for the realistic  renormalization procedure the 
following facts serves as the important guidelines.
1) The Wilsonian renormalization method \cite{wilson} 
and the surface theorem \cite{pol} which reveal that, only few 
operators play important role to
determine the behavior of the dynamic 
system at the low energy region,
 enable us to truncate the infinite divergence
tower up to a specified order and to consider the renormalization
of the effective Lagrangian order by order;
2) While in the dimensional regularization
method, although the quartic divergences exist in the 
loop calculation, it allows us to
express them to be proportional to the masses in the
theory, as done with the quadratic divergences.

The effective couplings of operators in the Lagrangian $L^{eff}$
form a parameter space of the effective
Lagrangian, and they effectively reflect the dynamics of the
underlying theories and the behavior of symmetry breaking.
Different underlying theories and ways of symmetry breaking
will fall into this effective parameter space as special points.
When the scale runs from the high energy region down
to the low energy region, a characteristic
curve in this parameter space is obtained. 
If we can measure or extract this curve from
the interpretation and extrapolation of experimental results, then it
would  help us to figure out
the possible underlying theories, as we attempt to do in speculating the
grand unification theories.

The magnitudes of these effective 
couplings are arbitrary. But generally speaking, 
the unitarity condition for the theoretical
prediction  puts a constraint on the magnitudes of the
effective couplings of these operators.
And the larger the ultraviolet cutoff 
$\Lambda_{UV}$ of the theory is, the smaller are the
magnitudes of the effective couplings of these operators \cite{yandu}.
However, here in order to keep the generality and universality
of the effective Lagrangian method, we do not make any 
assumption on the magnitude of these effective couplings 
and simply classify operators with the same mass 
dimensions as the same order.

For the $SU(2)$ case, the general 
effective $SU(2)$ Lagrangian $L^{eff}$ which is consistent with the
Lorentz spacetime symmetry, $SU(2)$ gauge symmetry, and the charge,
parity, and the combined CP symmetries, can be formulated as
\begin{subequations}
\bea
\label{effl}
{\cal L}_{eff} &=& {\cal L}_2 
+ {\cal L}_4 + {\cal L}_6 + \cdots \,,\label{efflag}\\
{\cal L}_2  &=& 
{v^2 \over 4} tr[V_{\mu} V^{\mu}]\,,\label{masst}\\
{\cal L}_4 &=& -{1\over 4 g^2} W_{\mn}^a W^{a\mn}
- i d_1 tr[W_{\mn} V^{\mu} V^{\nu}]
\nnb\\&&
+ d_2 tr[V_{\mu} V_{\nu}] tr[V^{\mu} V^{\nu}]
+ d_3 tr[V_{\mu} V^{\mu}] tr[V_{\nu} V^{\nu}]
\label{op4}\,,\\
{\cal L}_6&=&\cdots\,,
\eea
\end{subequations}
where the ${\cal L}_2$ and
${\cal L}_4$ represent the relevant 
and marginal operators in the Wilsonian
renormalization method, respectively. 
Henceforth for the simplicity,  we 
omit all the irrelevant operators in our consideration, 
i.e. higher dimensional operators greater 
than $O(p^4)$. The operators in the ${\cal L}_2$ and
${\cal L}_4$ also form the set of 
complete operators up to $O(p^4)$ in
the usual momentum counting rule.
And the higher dimension ( irrelevant ) 
operators greater than $O(p^4)$ order are contained in ${\cal L}_6$. 
The auxiliary variable $V_{\mu}$ is defined as
\bea
V_{\mu} = U^{\dgr} D_{\mu} U\,,\,\,
D_{\mu} U =\partial_{\mu} U - i W_{\mu} U\,,
\label{vdef}
\eea
to simplify the representation.
Due to the following relations of the $SU(2)$ gauge group
\bea
tr[T^a T^b T^c T^d]={1\over 8} (\delta^{ab} \delta^{cd}
+\delta^{ad} \delta^{bc}-\delta^{ac} \delta^{bd})\,,
\eea
the terms, like $tr[V_{\mu} V_{\nu} V^{\mu} V^{\nu}]$ 
and $tr[V_{\mu} V^{\mu} V_{\nu} V^{\nu}]$,
can be linearly composed by 
$tr[V_{\mu} V_{\nu}] tr[V^{\mu} V^{\nu}]$
and $tr[V_{\mu} V^{\mu}] tr[V_{\nu} V^{\nu}]$.
And since here we do not consider 
the operators  which break the charge,
or parity, or both symmetries,
therefore, the operators in Eq. (\ref{op4}) are complete
and linearly independent.
Before havng any  certain idea  on the 
magnitude of $d_i$, which does not possess any mass dimension, we can
classify them in the same class of the effective couplings 
 as that of $1/g^2$.
\section{The renormalization of the nonlinear SU(2) 
Higgs model by BFM}
\label{sec:su2hrenorm}
We formulate the partition 
function of the nonlinear $SU(2)$ Higgs 
model in Eq.(\ref{uglag}) in the BFM  as
\bea
{\cal Z}[{\overline W^s},{\overline H}] &= &
\int {\cal D} {\widehat W_{\mu}^a} {\cal D} {\widehat H} {\cal D} 
{\widetilde \xi^b} 
\det\left \{ \left  ({v + {\overline H} 
+  {\widehat H} \over v }\right  ) \delta(x-y) \right \}\nnb \\
&& \exp\left ( {\cal S}'[{\overline W^s},{\overline H};
{\widehat W},{\widehat H},{\widetilde \xi}] 
+ S^{\prime}_{GF} \right )\,,
\eea
where the classical field ${\overline W^s}$ is given
as
\bea
{\overline W}^s ={\overline U}^{\dgr} {\overline W} {\overline U}
+ i {\overline U}^{\dgr} \partial{\overline U}\,.
\label{strs}
\eea

From now on, in order to simplify expressions, 
we will abuse ${\overline W}$ to represent ${\overline W}^s$.
And ${\overline U}$ and ${\overline H}$ 
are the classical parts of the Goldstone
and Higgs fields, respectively. 
The quantum fields ${\widehat W_{\mu}^a}$,
${\widehat H}$, and ${\widetilde \xi^b}$ are the 
quantum parts of gauge, scalar, and Goldstone
fields. In the BFM \cite{bfm}, 
we can choose different gauges for the
classical and quantum fields, respectively. 
Although the classical fields might be restricted
by their classical EOM, 
the gauge fixing condition is still
needed in order to find one unique solution, 
as we do in the classical electrodynamics.
For the quantum fields, the gauge fixing 
terms would be used to eliminate the redundant gauge
degrees of freedom.

The action $S_{GF}^\prime$  contains the gauge fixing terms
of quantum gauge and Goldstone fields. 
Normally, for the sake of convenience,
we choose the Feynman-$^{\prime}$t 
hooft gauge for the $S^{\prime}_{GF}$, which is given as
\bea
S^{\prime}_{GF}&=&- {1 \over 2} \left ( D_{\mu} {\widehat  W}^{\mu,a} + (v
+ {\overline H} ) {\tilde \xi^a} \right )^2 \, .
\eea

By using the Euler-Lagrange equation for a field, 
the EOM of the classical 
gauge and Higgs bosons are determined as
\bea
D_{\mu}^{ab} {\overline W}^{b,\mu \nu} &=&
{1 \over 4} (v + {\overline H} )^2 {\overline W^{a,\nu}}\,,\\
\partial^2 {\overline H} &=& 
{1 \over 4} (v +{\overline H} ) {\overline W} \cdot {\overline W}
+ {\lambda \over 4} (v +{\overline H}) 
[ (v +{\overline H})^2- v^2]\,.
\label{heom}
\eea

The EOM of the classical gauge bosons yields the following
relation
\bea
\partial_{\mu} {\overline H} {\overline W^{a,\mu}} = 
-{ v + {\overline H} \over 2} \partial \cdot {\overline W^{a}}\,.
\label{veom}
\eea
This relation is very important to eliminate 
all those terms with $
\partial_{\mu} {\overline H} {\overline W^{a,\mu}}$.

In order to give the proper definition to the 
propagator of the quantum Goldstone fields as in contrary to that
given  given in Eq. (\ref{dao}), we
redefine the quantum Goldstone fields ${\widetilde \xi}$ as
\bea
{\widetilde \xi} \rightarrow 
{v \over v + {\overline H}} {\widehat \xi}\,.
\label{para}
\eea
Then the partition function changes to
\bea
{\cal Z}[{\overline W^s},{\overline H}] &=& 
\int {\cal D} {\widehat W_{\mu}^a} {\cal D} {\widehat H} {\cal D} 
{\widehat \xi^b} 
\det\left \{ \left  ({v + {\overline H}
+  {\widehat H} \over v + {\overline H} }\right  )
 \delta(x-y) \right \} \nnb\\
&&\exp\left ({\cal S}'[{\overline W},{\overline H};
{\widehat W},{\widehat H},{\widehat \xi}] + S_{GF}\right ) \,.
\label{mlag}
\eea
And the gauge fixing term $S_{GF}$ is also 
modified in terms of ${\widehat W_{\mu}^a}$
and ${\widehat \xi^b}$.
With this parameterization of 
the quantum Goldstone fields,
we can define the propagator of 
Goldstone boson properly in either the
coordinate or the momentum space. 
And  no quartic divergence appears
in the intermediate calculation steps. 
 The quartic divergences are collected in the Eq. (\ref{mlag}) as $det\{
\left  ({v + {\overline H}
+  {\widehat H} \over v + {\overline H} }\right  )
 \delta(x-y)
\}$.
 But if we 
parameterize the quantum Goldstone fields as given 
in \cite{Dittmaier} , it  becomes obscure to 
collect the quartic terms and verify
the renormalizability of the theory.

The  terms which are bilinear in quantum fields contribute to
the one loop effective Lagrangian and can be presented in a standard form.
However, these terms are obtained after a tedious calculation,
manipulating the partial integrals and using the antisymmetric
properties of SU(2) structure constants. 
\bea
{\cal L}_{quadratic\,\,terms} &=& -{1 \over 2} \left( 
{\widehat W^{a,\mu}} \Box_{\mu\nu}^{ab} {\widehat W^{b,\nu}} +
{\widehat \xi^{a}} \Box^{ab} {\widehat \xi^b} +
{\widehat H} \Box {\widehat H} + \right . \nnb \\ && \left .
{\widehat W^{a,\mu}} \lx_{\xi}^{\mu,ab} {\widehat \xi^b} + 
{\widehat \xi^a}     \rx_{\xi}^{\mu,ab} {\widehat W^{b,\mu}} +
{\widehat W^{a,\mu}} \lx_{H}^{\mu,a} {\widehat H} + 
\right .\nnb\\&& \left .
{\widehat H}     \rx_{H}^{\mu,a} {\widehat W^{a,\mu}} +
{\widehat \xi^a} \lx^{a} {\widehat H} + 
{\widehat H}     \rx^{a} {\widehat \xi^a}  
\right )
- {\bar c} \Box^{ab}_{{\bar c} c} c\,,
\eea
where the operators are defined as
\begin{subequations}
\bea
\Box_{\mu\nu}^{ab} &=& [-(D \cdot D)^{ab} 
+ m_W^2 \delta^{ab}] g^{\mu\nu} 
+ \sigma^{ab,\mu\nu}_W\,, \label{prop1}\\
\Box^{ab} &=& [- (d \cdot d)^{ab} 
+ m_W^2 \de^{ab} ] + \sigma^{ab}_{\xi} \,, \label{prop2}\\
\Box &=& [- \partial \cdot \partial + m_H^2] + \sigma_H \,,\\
\Box^{ab}_{{\bar c} c}&=& [-(D \cdot D)^{ab} + m_W^2 \delta^{ab}]
+ \sigma^{ab}_c\,, \label{prop3}\\
\lx_{\xi}^{\mu,ab} &=&g \left[-{i \over 2} ( v + {\overline H})
{\overline W^{ab}_{\mu,G}} 
+ \partial_{\mu} {\overline H} \delta^{ab} \right] \,,\label{prop4}\\
\rx_{\xi}^{\mu,ab} &=&g \left[ {i \over 2} ( v + {\overline H})
{\overline W^{ab}_{\mu,G}} 
+ \partial_{\mu} {\overline H} \delta^{ab} \right]\,,\label{prop5}\\
\lx_{H}^{\mu,a} &=& {g \over 2} 
( v + {\overline H}) {\overline W^{a}_{\mu}} \,,\label{prop6}\\
\rx_{H}^{\mu,a} &=&{g \over 2} 
( v + {\overline H}) {\overline W^{a}_{\mu}}\,,\label{prop7}\\
\lx^{a} &=& \lx^{\alpha,a}\partial^{\alpha} + \lx^{a}_0 \,,\\
\rx^{a} &=& \rx^{\alpha,b} d^{ba,\alpha} + \rx^{a}_0\,,\label{prop8}\\
\lx^{\alpha,a}&=&{\overline W^{a}_{\alpha}}\,,\label{prop9} \\
\lx^{a}_0&=&{\partial \cdot {\overline W^{a}} \over 2}\,,\\
\rx^{\alpha,b}&=&-{\overline W^{b}_{\alpha}}\,, \label{prop10}\\
\rx^{a}_0&=& -{\partial \cdot {\overline W^{a}} \over 2}\,,
\label{prop11}
\eea
\end{subequations}
where the ${\overline W^{ab}_{\mu,G}}$ is 
defined as ${\overline W^{ab}_{\mu,G}}
= {\overline W^{e}_{\mu}} \, (t^e)^{ab}$, 
and $t^e$ is the matrix of adjoint
representation the group. The covariant differential operator
$d_{\mu}$ of the Goldstone boson is defined as 
$d_{\mu}^{ab}=\partial_{\mu} \delta^{ab} - i \,g / 2\, {\overline W^{ab}_{\mu,G}}$,
while the covariant differential operator
$D_{\mu}$ of the gauge boson is defined as
$D_{\mu}^{ab}=\partial_{\mu} \delta^{ab} - i \,g \, {\overline W^{ab}_{\mu,G}}$.
In the $SU(2)$ case, 
$(t^e)^{ab} = i f^{aeb}$, and $f^{aeb}$
is the structure constant of the $SU(2)$ Lie algebra.
And the other quantities are defined as
\begin{subequations}
\bea
\sigma^{ab,\mu\nu}_W &=&\left[ {g^2 \over 4} (v + {\overline H} )^2
 - {g^2 \over 4} v^2\right] \de^{ab} g^{\mu\nu}
+ 2 i W^{ab,\mu\nu}\,,\label{sigmaW}\\
\sigma^{ab}_{\xi} &=& \left[ {g^2 \over 4} 
(v + {\overline H} )^2 - {g^2 \over 4} v^2 \right]\,\delta^{ab}
- {1 \over 4} {\overline W^{ac}_G} \cdot {\overline W^{cb}_G} 
+ {\partial^2 {\overline H} \over v 
+ {\overline H}} \delta^{ab}\,,\label{sigmaG}\\
\sigma_H &=& {1 \over 4} {\overline W} \cdot {\overline W} +
{3 \lambda \over 2} v {\overline H} 
+ {3 \lambda \over 4} {\overline H}^2\,,\label{sigmaH}\\
\sigma_c^{ab} &= & \left[{g^2 \over 4} (v + {\overline H} )^2
 - {g^2 \over 4} v^2 \right]\,\delta^{ab}\,.
\label{sigmaGH}
\eea
\end{subequations}
Following the diagonalisation method prescribed in \cite{Dittmaier}, 
we  perform the Gaussian integral 
over quantum fields given by ${\cal L}_{quadratic\,  terms}$ in
Eq. (\ref{mlag}). Hence we calculate  the one loop effective
generating functional 
$\Gamma^{eff}_{1-loop}$, which is   relevant up to $O(p^4)$ 
\bea
\Gamma^{eff}_{1-loop} &=& - { 1\over 2 }\,\, \,  Tr \left[
 \ln \Box_W + \ln \Box_{\xi} + \ln \Box_H - 2 \ln \Box_c 
\ssep -\rx_{\xi} \Box_{W}^{-1} \lx_{\xi} \Box_{\xi}^{-1}
-\rx_{H} \Box_{W}^{-1} \lx_{H} \Box_{H}^{-1}
-\rx \Box_{\xi}^{-1} \lx \Box_{H}^{-1}
\ssep + \rx_{H} \Box_{W}^{-1}   \lx_{\xi} 
\Box_{\xi}^{-1} \lx \Box_{H}^{-1}
+ \rx \Box_{\xi}^{-1} \rx_{\xi} \Box_{W}^{-1} \lx_{H} \Box_{H}^{-1}
\ssep -{1 \over 2} \rx_{\xi} \Box_{W}^{-1} \lx_{\xi} 
\Box_{\xi}^{-1} \rx_{\xi} \Box_{W}^{-1} \lx_{\xi} \Box_{\xi}^{-1}
-{1 \over 2} \rx_H \Box_{W}^{-1} \lx_H 
\Box_{H}^{-1} \rx_H \Box_{W}^{-1} \lx_H \Box_{H}^{-1}
\ssep -{1 \over 2} \rx \Box_{\xi}^{-1} \lx 
\Box_{H}^{-1} \rx \Box_{\xi}^{-1} \lx \Box_{H}^{-1}
- \rx_H \Box_{W}^{-1} \lx_{\xi} \Box_{\xi}^{-1} 
\rx_{\xi} \Box_{W}^{-1} \lx_H \Box_{H}^{-1}
\ssep - \rx \Box_{\xi}^{-1} \lx_{\xi} 
\Box_{W}^{-1} \rx_{\xi} \Box_{\xi}^{-1} \lx \Box_{H}^{-1}
- \rx \Box_{\xi}^{-1} \lx \Box_{H}^{-1} 
\rx_H \Box_{W}^{-1} \lx_H \Box_{H}^{-1} \ssep
+ \cdots
\right]\,.
\label{1llag}
\eea
Here the  trace is over the points of the space-time,
Lorentz and group indices.
All these terms in the $\Gamma^{eff}_{1-loop}$ 
can be expressed by a set of 
Feynman diagrams at the one-loop level. 
Although it may not be an one-to-one correspondence,
but it guarantees the gauge 
invariance in each step of the calculation. 
\begin{figure}[b]
\vskip -3.50 cm
\centerline{
\epsfxsize= 16.0 cm\epsfysize=18.0cm
 \epsfbox{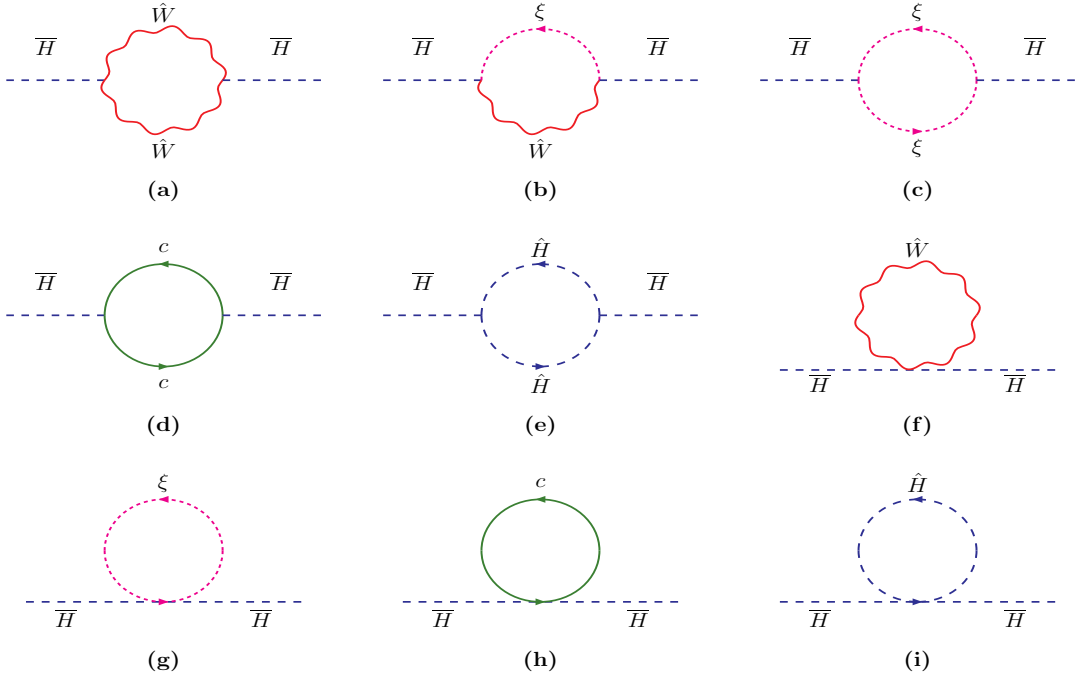}}
\vskip -5 cm
\caption{\it Diagrams  contributing to the
$\delta m_H^2$ in the coordinate space calculation in ${\overline{MS} }$ scheme. In the decoupling limit 
only Figs. (c),(e) and (i) contributes to $\delta m^2_H$}
\label{fig1}
\end{figure}
\vskip 0.2 cm
By using the algebraic calculation we can skip the correspondence with  Feynman
diagrams at each step of calculation and just extract all 
the relevant terms that we are interested in, but we believe that it is always  helpful to 
present linkage to the Feynman diagrams so as to make the comparison
convenient and easy. 

Considering  the importance of the one loop correction to the Higgs
boson mass,  we only display the Feynman diagrams which contribute to
this correction  as shown in Fig. (\ref{fig1}). 

Next, we evaluate the one-loop effective generating functional  $\Gamma^{eff}_{1-loop}$
given in Eq. (\ref{1llag}) in the coordinate space 
by using the heat kernel method 
and the covariant short distance expansion.
The basics of the heat kernel method and the short distance expansion
is provided in  appendix \ref{app:heatkernel}.
The contributions of each term in $\Gamma^{eff}_{1-loop}$
are listed in appendix \ref{app:div}.
With these formalism, the divergent terms in the one-loop
effective Lagrangian are extracted and can be expressed as:
\bea
{\cal L}^{eff}_{1-loop}(divergent) &=&
- {1 \over 16 \pi^2} {2 \over \bar \epsilon} \left \{
-{43 \over 6} {1\over 4 }
 {\overline W_{\mn}^a} {\overline W^{a\mn}}
-{3 \over 16} {1 \over 4} (\WW \cdot \WW)^2
\ssep+ {3 \over 8} {\partial^2 \HH \over v+\HH} \WW \cdot \WW
-{3 \over 4} {(\partial^2 \HH)^2 \over (v+\HH)^2}
\ssep -{9  \over 32 } g^2 (v+ \HH)^2 \WW \cdot \WW
- {3 \over 8} g^2 (v + \HH) \partial^2 \HH
\ssep -{3 \over 2} g^2 \partial_{\mu}\HH \partial^{\mu} \HH
-{1\over 64} \lambda^2 v^4 + {3 \over 32} \lambda^2 (v+\HH)^2 v^2
\ssep -{9 \over 64} (g^4 + \lambda^2) (v + \HH)^4
\right \}\,,
\label{divlag}
\eea
where the $2/\bar\epsilon$ is defined as 
$2/\epsilon - ln (4 \pi) + \gamma_E $.
Here we find there are extra divergent structures,
like $(\WW \cdot \WW)^2$, $\partial^2 \HH \WW \cdot \WW$,
and $(\partial^2 \HH)^2$, etc.
\par We now use the EOM for Higgs and gauge bosons to eliminate these
 extra divergent structures and get
\bea
\label{1ldiv}
{\cal L}^{eff}_{1-loop}(divergent) &=&
- {1 \over 16 \pi^2} {2 \over \bar \epsilon} \left \{
-{43 \over 6} {1\over 4 }
 {\overline W_{\mn}^a} {\overline W^{a\mn}}
- {3 g^2 \over 2} \partial {\overline H} 
\cdot \partial {\overline H}
\ssep - {3 g^2 \over 8 } {\overline W} 
\cdot {\overline W} (v + {\overline H} )^2
+ {3 \lambda \over 32} ( \lambda + 2 g^2) v^2 
(v + {\overline H} )^2
\ssep - {3 \over 64} (3 g^4 + 2 \lambda g^2 
+ 4 \lambda^2) (v + {\overline H} )^4
- {\lambda^2 \over 16} v^4
\right \}\,.
\label{div}
\eea
With the successful elimination of the extra divergent terms we
re-establish the renormalizability of the theory which by itself
provides the consistency check of our calculational strategy.
There are couple of points worthy of remarks regarding the calculation of
these divergent terms.
\begin{enumerate}
\item As a result of usage of  
EOM of the gauge bosons given in Eq. (\ref{veom}), 
there is no term $(\partial \cdot {\overline W})^2$ 
appearing in the divergent 
part given in Eq. (\ref{div}) and  hence 
it is no longer necessary to define the renormalization 
constant of the gauge parameter of the gauge fixing term of the 
classical gauge field. 
\item The use of EOM has its own ambiguities, {\it e.g.}
consider the term, $(v + \HH) \partial^2 \HH$. Here  one can use
 the EOM of classical Higgs field
either to eliminate $\partial^2 \HH$ or one can also replace the $( v
+ \HH ) \partial^2 \HH$ integral 
by $\partial_{\mu} \HH \partial^{\mu} \HH + v \partial^2 H$.
This ambiguity can be resolved by considering the  fact that
 the operators 
$\partial {\overline H} \cdot \partial {\overline H}/2$ 
and ${\overline W} \cdot {\overline W} (v + {\overline H} )^2/8$, in
Eq. (\ref{nonlinear}),   essentially stems  from 
$(D_\mu \phi)^{\dagger} \cdot D^{\mu} \phi$ in the linear form
given in Eq. (\ref{linear}).
Therefore,  the renormalization constants of these two operators
should always be same and this can be only achieved by using the EOM of Higgs
boson.
\item We have performed our calculation using  the Feynman-$ ^\prime $t Hooft
gauge $a=1$ and  ${\overline{MS}}$ renormalization scheme in the
coordinate space formalism.   $\de m_H^2$, in this
formalism,     at the decoupling limit can be
expressed in terms of the scalar $B_0$  integrals as demonstrated in the appendix \ref{app:B0}.
\bea
\de m_H^2 &=& {g^2 \over 16 \pi^2}{3 m_H^2 \over 8 m_W^2} 
\left \{
m_H^2 B_0(0, a \, m_W, a \, m_W) + 3 m_H^2 B_0(0, m_H,m_H)
+ A_0(m_H)
\right \}\,,\label{dmhcs}
\eea
Interpreting the Eq.(\ref{dmhcs}) in the diagrammatic language, we find
that at the decoupling limit, Fig.  \ref{fig1}(b)  contributes 
the term $B_0(0, m_W, m_W)$, Fig. \ref{fig1}(e)  contributes 
the term $m_H^2 B_0(0, m_H,m_H)$ and Fig. \ref{fig1}(i)  contributes 
the term $A_0(m_H)$.

\par Performing the same calculation  in the momentum space with
Landau  gauge $a=0$ and  On-Shell
renormalization scheme at the decoupling limit, yields
\bea
(\de m_H^2)^{On} &=& {g^2 \over 16 \pi^2}{3 (m_H^2)^{On} \over 8 m_W^2} 
\left \{
(m_H^2)^{On} Re(B_0(p^2|_{p^2=(m_H^2)^{On}}, a\, m_W, a\, m_W)) 
\ssep + 3 (m_H^2)^{On} B_0(p^2|_{p^2=m_H^2}, m_H^{On},m_H^{On})
+ A_0(m_H^{On})
\right \}
\,.\label{dmhms}
\eea
 which is in 
complete agreement with the reference \cite{Dittmaier}.

This would mean, the physical measurable quantities
 of the reference \cite{Dittmaier} can be
retrieved at the decoupling limit, 
provided we make a proper correspondence of the coordinate
space calculation in Feynman-$ ^\prime $t Hooft 
gauge and  ${\overline{MS}}$ renormalization scheme to the 
momentum space calculation in Landau  gauge and  On Shell
renormalization scheme along with the input 
of the relation 
\bea
(m_H^2)^{On}= m_H^2 + \Sigma(p^2,m_i^2,m_j^2)|_{p^2=(m_H^2)^{On}},
\label{dmhrltn}
\eea
\end{enumerate}
where $\Sigma$ is the self energy corrections to the two point
function of Higgs boson, the $p$ is the external momentum, 
and $m_i$, $m_j$ are the masses of particles in the loop.

\par We proceed to construct explicitly the counter-terms  from Eq. (\ref{div}).  
The one-loop counter Lagrangian includes  the terms which are
linear in $\de g$, $\de v$, $\de \mu^2$ and $\de \lambda$,
and can be expressed as
\bea
\de {\cal L'} &=& {\de g \over 2 g^3} W_{\mn}^a W^{a \mn}
	 - \left \{ (2 \de v - \de Z_H v) + \de Z_H (v + \HH)
\right \} {(v + \HH) \over 4 } tr[(D U)^{\dagger} \cdot (D U)] \nnb\\    
&&	- {1 \over 2} \de Z_{H} \partial \HH \cdot \partial \HH
	-{ 1 \over 2} (-2 \de v + \de Z_H v) \mu^2 ( v + \HH)
\jsep	+ {1 \over 2} \left ( \de \mu^2  + \mu^2 \de Z_H \right ) (v + \HH)^2
	- {\lambda \over 8} \left ( 2 \de v - \de Z_H v\right ) (v + \HH)^3
\jsep 	- {\de \lambda + 2 \de Z_H \lambda \over 16} (v+\HH)^4\,,
\label{ctlag}
\eea
where $Z_{H}$ is the renormalization constant
of Higgs field.

At the tree level, we can replace the the set of
parameters of the theory such as $v, \mu^2, \lambda $   in terms of 
the another set of of parameters namely the tadpole parameter $t$, the 
gauge boson's  mass $m_W$, 
and the Higgs boson's  mass $ m_H $ by the following equations 
\begin{subequations}
\bea
t &=& \mu^2 v - \frac{\lambda}{4} v^3\,, \label{pramsetA1}\\ 
m_W^2 &=& {1 \over 4} g^2 v^2\,, \label{pramsetA2}\\
m_H^2 &=& {1 \over 2} \lambda v^2\,\label{pramsetA3}.
\eea
\end{subequations}
And the corresponding relations of the counter terms
between these two sets of parameters are given as
\begin{subequations}
\bea
\frac{\de v}{v} &=&
\frac{1}{2} \frac{ \de m_W^2}{m_W^2}-\frac{\de g}{g},\label{cctm01} \\
\frac{\de \mu^2}{\mu^2} &=&
\frac{\de m_H^2}{m_H^2}+\frac{3g}{2 m_W m_H^2}\de t, \label{cctm02}\\
\frac{\de\lambda}{\lambda} &=&
\frac{\de m_H^2}{m_H^2}-\frac{\de m_W^2 }{ m_W^2 }+2\frac{\de g}{g}
+\frac{g}{2 m_W m_H^2}\de t\,.
\label{cctm03}
\eea
\end{subequations}
Requiring combination of Eq. (\ref{div}) and (\ref{ctlag}) to vanish, 
 the  counter terms at one loop level satisfy 
\begin{subequations}
\bea
{\de g  \over g}&=& g^2 {1 \over 16 \pi^2} {1 \over \bar \epsilon}
\left \{-{43 \over 12} \right \}\,,\label{ct1loop1} \\
{\de v  \over v}&=& {1 \over 16 \pi^2} {1 \over \bar \epsilon}
\left \{
 {3 \over 2} g^2
\right \}\,,\label{ct1loop2}\\
{\de \mu^2 \over \mu^2} &=& {1 \over 16 \pi^2} {1 \over \bar \epsilon}
\left \{ {3 \over 4} ( \lambda - 2 g^2)
\right \}\,,\label{ct1loop3}\\
{\de \lambda \over \lambda} &=& {1 \over 16 \pi^2} {1 \over \bar \epsilon}
\left \{ {3 \over 4} (3 {g^4 \over \lambda} - 6 g^2 
+ 4 \lambda )\right \}\,. \label{ct1loop4}
\eea
\end{subequations}
In addition to Eqs. (\ref{ct1loop1}-\ref{ct1loop4}), we also  have $ \de Z_H = 2 \de v/v$, which
guarantees that the vanishing of the divergences in the linear and trilinear
interaction of the Higgs potential.

Then from Eqs. (\ref{cctm01}-\ref{cctm03}), the terms $\de m_W^2$, $\de m_H^2$ and $\de
t$ are determined as
\begin{subequations}
\bea
\de m_W^2 &=&
2 m_W^2 \left \{ \frac{\de v}{v} + \frac{\de g}{g} \right \} \nnb\\
&=& m_W^2 {1 \over 16 \pi^2} {1 \over \bar \epsilon}
g^2 \left \{  -{25 \over 6} \right \}\,,\label{cctm11}\\
\de m_H^2 &=&
{m_H^2 \over 2} \left \{ 
3  \frac{\de\lambda}{\lambda}
+ 6 \frac{\de v}{v}
- \frac{\de \mu^2}{\mu^2}
\right \} \nnb \\
&=& m_H^2 {1 \over 16 \pi^2} {1 \over \bar \epsilon}
\left \{ 
-3 g^2 + {27 \over 4} {g^4 \over \lambda} + {33 \over 4} \lambda
\right \}\,,\label{cctm12} \\
\de t &=& {\lambda \mu^2 \over 4} 
\left \{ \frac{\de \mu^2}{\mu^2} - \frac{\de\lambda}{\lambda}
-2 {\de v \over v} \right \} \nnb\\
&=& {9 \mu^2 \over 16} {1 \over 16 \pi^2} {1 \over \bar \epsilon}
\left \{ 
g^4 + \lambda^2
\right\}
\,.
\label{cctm13}
\eea
\end{subequations}
With those divergent terms given 
in Eq. (\ref{1ldiv}) and the  relations given in Eqs. (\ref{cctm11}-\ref{cctm13}), 
it is straightforward to construct 
the renormalization constants in 
the modified minimal subtraction scheme
and derive the corresponding running
behavior of those parameters in the theory. 
However,  the masses of gauge and
Higgs boson given here are not the physical 
parameters measured in the experiments.

\section{The complete anomalous couplings up to one-loop level}
\label{sec:com1loop}
From the one-loop effective Lagrangian given in Eq. (\ref{1llag}) and
the technique in the appendix \ref{app:div},
we can also extract those finite
terms which contribute to the trilinear 
and quartic anomalous interaction vertices.
The finite part of the one-loop effective Lagrangian
can be formulated as
\bea
{\cal L}^{eff,full}_{1-loop}(finite) &=& 
  {\de t} \HH
+ {\de m_W^2 \over 2} W \cdot W
+ {\de m_H^2 \over 2} \HH^2
+ {\de \lambda_3} \HH^3
\jsep + {\de \lambda \over 16} \HH^4
+ {\de g \over 2 g^3} W_{\mn} W^{\mn}
+ d_1^{full} tr[W_{\mn} V^{\mu} V^{\nu}]
\jsep
+ d_2^{full} tr[V_{\mu} V_{\nu}] tr[V^{\mu} V^{\nu}]
+ d_3^{full} tr[V_{\mu} V^{\mu}] tr[V_{\nu} V^{\nu}]
+ \cdots \,.
\eea
The dots represent higher order terms,
like $\bigg(\HH / v\bigg)\,\, W_{\mn} W^{\mn}$, {\it etc}.

The complete one-loop anomalous 
couplings of the vector boson sector
in the renormalizable SU(2) theory
can be divided into two parts:
one is purely 
from the contribution 
of the gauge and the Goldstone bosons, the other
is from the Higgs boson and its mixing 
with gauge and Goldstone bosons, {\it i.e.}
$d_i^{full} = d_i^{c,full} + d_i^{h,full}$. 
Although both the $d_i^{c,full}$ and
$d_i^{h,full}$ are renormalization scale 
dependent, the combination of them
indeed yield a renormalization scale independent results.
The underlying reason for this is that the $SU(2)$ Higgs model is
a renormalizable theory.

If we set the renormalization scale $\mu_{E}=m_W$, then $d_i^c$ become constant part, which is given as
\begin{subequations}
\bea
d_1^{c,full} &=&{1 \over 16 \pi^2} \left( -{11 \over 6}\right ) \,,\label{fulld1}\\
d_2^{c,full} &=&{1 \over 16 \pi^2} \left( {1 \over 2} \right )\,,\label{fulld2}\\
d_3^{c,full} &=&{1 \over 16 \pi^2} \left( - {1 \over 6} \right )\,.
\label{fulld3}
\eea
\end{subequations}
The Higgs  boson dependent terms of anomalous couplings are given as
\begin{subequations}
\bea
d_1^{h,full} &=&{1 \over 16 \pi^2} \left\{
\frac{-197 + 154\,r - 17\,r^2}{72\,{\left( -1 + r \right) }^2} + 
  \frac{r\,\left( 36 - 27\,r + r^2 \right) \, \ln (r)}
   {12\,{\left( -1 + r \right) }^3}
\right \}\,, \label{d1h}\\
d_2^{h,full} &=&{1 \over 16 \pi^2} \left\{
\frac{-71 + 208\,r - 17\,r^2}{72\,{\left( -1 + r \right) }^2}  + 
  \frac{r^2 \,\left( -21 + r  \right) \,\ln (r)}
   {12\,{\left( -1 + r \right) }^3}
\right \}\,,\label{d2h}\\
d_3^{h,full} &=&{1 \over 16 \pi^2} \left\{
\frac{13 - 374\,r + 13\,r^2}{144\,{\left( -1 + r \right) }^2}  +
  \frac{\left( 27 + 3\,r + 87\,r^2 - r^3 \right) \,\ln (r)}
   {48\,{\left( -1 + r \right) }^3}
\right \}\,,
\label{d3h}
\eea
\end{subequations}
where $r=m_H^2/m_W^2$. 

The anomalous couplings given in Eqs. (\ref{d1h}-\ref{d3h}) might be infrared
singular at their appearance when $r$ approaches to $1$, 
but we find that this is not true, actually the infrared singularities
from the non-log and log parts just cancel exactly with each other.
In the limit with $r$ approaches to $1$, we have
\begin{subequations}
\bea
d_1^{h,full} &=& {1 \over 16 \pi^2} \left (- {4 \over 3} \right ) \,, \label{d1hr1}\\
d_2^{h,full} &=& {1 \over 16 \pi^2} \left (- {2 \over 3} \right ), \label{d2hr1}\\
d_3^{h,full} &=& {1 \over 16 \pi^2} \left ( {5 \over 6} \right )\,. \label{d3hr1}
\eea
\end{subequations}

In the limit with $r \rightarrow \infty$, 
the non-decoupling parts
of these $d^h_i$ are given as
\begin{subequations}
\bea
d^{h,full}_1 &=&{1 \over 16 \pi^2} \left\{ 
- {17 \over 72} + {1 \over 12} \ln (r) \right \}\,,\label{mwdcpl1}\\
d^{h,full}_2 &=&{1 \over 16 \pi^2} \left\{ 
- {17 \over 72} + {1 \over 12} \ln (r) \right \}\,,\label{mwdcpl2}\\
d^{h,full}_3 &=&{1 \over 16 \pi^2} \left\{ 
  {13 \over 144} - {1 \over 48} \ln (r) \right \}\,.
\label{mwdcpl3}
\eea
\end{subequations}
In order to derive the one-loop matching conditions,
we should also perform the one-loop calculation for
the non-renormalizable 
effective chiral $SU(2)$ theory. 
After performing the path integral
as prescribed in reference \cite{yandu} and  appendix \ref{app:rge},
we  construct the counter terms to eliminate the divergences.
Then we  formulate the finite one-loop effective Lagrangian
as
\bea
{\cal L}^{eff,chl}_{1-loop}(finite) &=& 
{\de m_W^2 \over 2} W \cdot W
+ {\de g \over 2 g^3} W_{\mn} W^{\mn}
+ d_1^{eff} tr[W_{\mn} V^{\mu} V^{\nu}]
\jsep
+ d_2^{eff} tr[V_{\mu} V_{\nu}] tr[V^{\mu} V^{\nu}]
+ d_3^{eff} tr[V_{\mu} V^{\mu}] tr[V_{\nu} V^{\nu}]
+ \cdots \,. \label{finchl1loop}
\eea

As in the full theory, $d_i^{eff}$ can also be decomposed into
two parts: the part from the contribution of Goldstone
and vector bosons, $d_i^{c,eff}$, and the part dependent on
the Higgs boson's mass, $d_i^{h,eff}$. We perform the similar
computational steps with the SU(2) chiral Lagrangian given in 
Eq. (\ref{effl}) to extract the anomalous couplings $d_i^{eff}$.
The contribution of gauge and Goldstone bosons
is  given as (here we have substituted
the renormalization scale $\mu_{E}=m_H$):
\begin{subequations}
\bea
d_1^{c,eff} &=&{1 \over 16 \pi^2} \left(- {11 \over 6} + {1 \over 12 } \ln (r) \right )\,, \label{effd1}\\
d_2^{c,eff} &=&{1 \over 16 \pi^2} \left( {1 \over 2} + {1 \over 12 } \ln (r) \right )\,,  \label{effd2}\\
d_3^{c,eff} &=&{1 \over 16 \pi^2} \left( {1 \over 6} + {1 \over 24 } \ln (r) \right )\,.
\label{effd3}
\eea
\end{subequations}
In order to compute the effective Lagrangian (\ref{finchl1loop}) up to the
one-loop level, we require the tree level matching conditions
as input. According to the tree-level matching conditions,
which are presented later  in Eqs. (\ref{tree1}-\ref{tree3}), we know 
that the terms dependent on the Higgs boson's mass
are those which are dependent on $d_3^{tree}$, so we only retain terms
proportional to $d_3^{tree}$ and its powers. Then we have
the one-loop anomalous couplings which are given as
\begin{subequations}
\bea
d_1^{h,eff} &=&{1 \over 16 \pi^2} \left(
d_3^{tree} g^2 -5 d_3^{tree} g^2 \ln(r) \right )\,, \label{d1eff1-loop}\\
d_2^{h,eff} &=&{1 \over 16 \pi^2} \left( 
{3 \over 2} (d_3^{tree})^2 g^4 + (d_3^{tree})^2  g^4 \ln(r)
- d_3^{tree} g^2 \ln(r) 
\right )\,,  \label{d2eff1-loop}\\
d_3^{h,eff} &=&{1 \over 16 \pi^2} \left(
-{39 \over 4} (d_3^{tree})^2 g^4 + 6 d_3^{tree} g^2
+ {25 \over 2 } (d_3^{tree})^2 g^4 \ln(r)
- 6 d_3^{tree} g^2 \ln(r) 
\right )\,.
\label{d3eff1-loop}\eea
\end{subequations}

\section{The tree level and the one-loop matching conditions}
\label{sec:treematch}
According to the standard matching procedure specified in \cite{Georgi},
we should match the full theory and 
the effective theory  order by order.
The effective couplings are organized in term of loops as
\bea
d_i &=& d_i^{tree} + d_i^{1-loop} + ...\,.
\eea
To determine the higher order of 
couplings, we need to know the lower
order ones.

In the $SU(2)$ Higgs model case, at the tree level, it suffices to
integrate out the Higgs
boson, using the EOM by 
assuming that $\partial^2 {\overline H}$ is negligibly  small,
which then expresses the Higgs boson in terms of the low energy 
dynamical degrees of freedom
and can be formulated as
\bea
\label{eom}
{\overline H} &=& {v \over 2 m_H^2} (D U)^{\dagger}\cdot(D U) + \cdots\,,
\eea
 The omitted
terms contain at least four covariant partials and belong
to the higher order operators.

To set up the matching conditions at the tree level we first 
 use Eq. (\ref{eom})
to eliminate the Higgs field in the modified Lagrangian given in  Eq. (\ref{hml}), at the 
matching scale ( which is always taken at the
scalar mass $\mu_E=m_H$ ). Then we compare the terms of this resulting
Lagrangian with the effective Lagrangian given in Eq. (\ref{op4}).
Thus the matching conditions for the effective couplings
at the tree level are determined as
\begin{subequations}
\bea
d_1^{tree}(m_H) &=&\,\, 0\,, \label{tree1}\\
d_2^{tree}(m_H)& =&\,\, 0\,, \label{tree2}\\
d_3^{tree}(m_H) &=& {v^2 \over 8 m_H^2}
={m_W^2 \over 2 g^2 m_H^2}\,,\,
\cdots\,.
\label{tree3}
\eea
\end{subequations}
In its decoupling limit $m_H 
\rightarrow \infty$ ($\lambda \rightarrow \infty$),
all these three effective couplings vanish. 
Generally, for certain  theoretical reasons (say, the validity of
perturbation theory )
$\lambda$ does not tend to 
$\infty$, and is usually considered to be of the $O(1)$, as
we realize in the case of standard model. So, $d_3^{tree}$ can be quite large compared with
other anomalous couplings.
However, in technicolor theories, all of these anomalous couplings might
be quite large.

To derive the matching conditions at one-loop level, at the first step,
both the calculations in the effective as well as in the  full theory 
are to be considered up to the one-loop order. As standard in perturbation theory,
we would require the tree level results 
of $d_i^{tree}(m_H)$ as inputs to compute the higher order
results $d_i^{1-loop}(m_H)$. Then, at the second step, while
 matching the one-loop effective Lagrangian of these 
two theories, we eliminate  Higgs field using EOM given in (\ref{eom}) 
as specified in Eq. (\ref{mtce}). Following these two steps
and using the one-loop anomalous couplings of the full theory
in Eqs. (\ref{fulld1}-\ref{fulld3}) and (\ref{d1h}-\ref{d3h}) and those in the effective theory in
Eqs. (\ref{effd1}-\ref{effd3}) and (\ref{d1eff1-loop}-\ref{d3eff1-loop}),
we arrive at the following matching conditions.
\begin{subequations}
\bea
d_1^{1-loop}(m_H) &=& {1 \over 16 \pi^2} \left \{
\frac{-\left( 36 + 125\,r - 118\,r^2 + 17\,r^3 \right) }
   {72\,{\left( -1 + r \right) }^2\,r} \ssep - 
  \frac{\left( 30 - 91\,r + 57\,r^2 - 6 r^3 \right) \,\ln (r)}
   {12\,{\left( -1 + r \right) }^3\,r}
\right \}\,,\, \label{d11loop}\\
d_2^{1-loop}(m_H) &=& {1 \over 16 \pi^2} \left\{
- \frac{27 - 54\,r + 98\,r^2 - 208\,r^3 + 17\,r^4}
   {72\,{\left( -1 + r \right) }^2\,r^2} \ssep -
  \frac{\left( -3 + 15\,r - 28 \,r^2 
+ 24\,r^3 + 12 \,r^4 \right) \,\ln (r)}
   {12\,{\left( -1 + r \right) }^3\,r^2}
\right \}\,,\, \label{d21loop}\\
d_3^{1-loop}(m_H) &=& {1 \over 16 \pi^2}\left\{
\frac{1458 - 3150\,r + 1763\,r^2 - 1018\,r^3 + 107\,r^4}
   {288\,{\left( -1 + r \right) }^2\,r^2} \ssep - 
  \frac{\left( 120 - 318\,r + 145\,r^2 + 69\,r^3 - 156\,r^4 \right) \,
     \ln (r)}{48\,{\left( -1 + r \right) }^3\,r^2}
\right \} \,.
\label{d31loop}
\eea
\end{subequations}
Here it is worth mentioning 
that two operators, $H^2$ 
and $H {\overline W} \cdot {\overline W}$,
in the one-loop effective generating functional of
the full theory also contributes to $d_3^{1-loop}(m_H)$.

In the limit with $r \rightarrow \infty$, the non-decoupling parts
of these $d^{1-loop}_i$ are given as
\begin{subequations}
\bea
d_1^{1-loop}(m_H) &=&{1 \over 16 \pi^2} \left(
- {17 \over 72} \right) \,, \label{mhdcpl1}\\
d_2^{1-loop}(m_H)& =&{1 \over 16 \pi^2} \left(
- {17 \over 72} \right) \,, \label{mhdcpl2}\\
d_3^{1-loop}(m_H) &=&{1 \over 16 \pi^2} \left (
 {107 \over 288} \right )\,.
\label{mhdcpl3}
\eea
\end{subequations}
From  Eqs. (\ref{mhdcpl1}-\ref{mhdcpl3}), we find that at the decoupling limit, the matching conditions
shred off their dependencies on  the non-decoupling 
logarithms. 
 These constants, when compared with
those obtained in \cite{Dittmaier}, are same for the first two
 anomalous couplings while it appears to be different for the 
 $d_3^{1-loop}(m_H)$. This reason for this difference can be traced back
 in the difference arising in
$\de m^2_H$ given in Eq.(\ref{dmhcs}) and Eq. (\ref{dmhms}) and their
 evaluation  as given in appendix \ref{app:B0}. The $p^2$ dependence
in the scalar $B_0$ integrals  affects the finite part of
$\de m_H^2$, and manifests its bearing in the evaluation of  $d_3^{1-loop}$ when
the Higgs field is integrated out. 
Although $d_3^{1-loop}$ given here is different from that in
\cite{Dittmaier}, the total $d_3=d_3^{tree} + d_3^{1-loop}$, which is
the physical and detectable quantity in experiments, 
is just the same for both renormalization schemes when the
relation of the Higgs boson's mass between them given in
Eq. (\ref{dmhrltn}) is used. 

\section{Comparison with the 
renormalization group equation method}
\label{sec:compare}
The RGE of an effective theory is one of 
the its basic ingredients, and its function is
to sum over quantum corrections 
in leading log (one-loop RGEs), next leading log (two-loop
RGEs), and so on. For instance, the one-loop RGEs 
can sum over all leading log of all loop diagrams.
 However, it differs
from the direct loop calculations, where the radiative
contribution are computed and organized by loops.

\par In order to check the reliability of the effective field
theory method, it is constructive to compare the 
predictions of the one-loop
direct calculation and those of the one-loop RGEs.
The one-loop RGEs of the chiral effective SU(2) 
Lagrangian can be derived by computing the
one-loop irreducible vertex generating functional,
and has been computed  in
the reference \cite{yandu}. 
Appendix \ref{app:rge} outlines this calculation with the corrections
to reference \cite{yandu}. 
These corrected RGEs can be tabulated as
\begin{subequations}
\bea
{d g^2\over dt} &=&{g^4\over 8 \pi^2} \left[
-\frac{29}{4}   - \frac{20 \don g^2}{3} - 
  \frac{23 {\don}^2 g^4}{24}
\right]\,, \label{rgeeg}\\
{d v\over dt}   &=&{ v \over 8 \pi^2} \left [
\frac{3 g^2}{2}+ \left( {5} \don - {10} \dtw - 
     \frac{35 \dth}{2} \right)  g^4 + 
  \frac{13 {\don}^2 g^6}{8}
 \right ]\,,\label{rgeev}\\
{d d_1\over dt} &=&{ 1 \over 8 \pi^2} \left \{
- \frac{1}{12}+
  \left( - \frac{26 \don}{3} - \frac{5 \dtw}{2} + 
     5 \dth \right)  g^2
 \right. \nnb\\ && \left.- 
  \frac{109 {\don}^2 g^4}{12} - 
  \frac{19 {\don}^3 g^6}{12}
 \right \}\,,  \label{rgeed1}\\
{d d_2 \over dt}&=&{ 1 \over 8 \pi^2} \left \{
- \frac{1}{12}   + 
  \left( \frac{ \don}{2} + 3 \dtw + 
      \dth \right)  g^2 \ssep + 
  \left [ \frac{87 {\don}^2}{32} - 6 {\dtw}^2 -
     5 \dtw \dth - {\dth}^2 + 
     \don \left( 5 \dtw - 
        4 \dth \right)  \right ]  g^4 \ssep + 
  \left[ - {\don}^3 + 
     {\don}^2 \left( -{5\over 4} \dtw - 
        \frac{5 \dth}{4} \right)  \right ]  g^6 \ssep
  - \frac{43 {\don}^4 g^8}{24}
 \right \}\,,\label{rgeed2}\\
{d d_3\over dt} &=&{ 1 \over 8 \pi^2} \left\{
- \frac{1}{24} + 
  \left( \frac{5 \dtw}{2} + 6  
      \dth \right)  g^2\ssep + 
  \left[ - \frac{155 {\don}^2}{32} -
     \frac{9 {\dtw}^2}{4} + 
     \don \left( 10 \dtw + 28
        \dth \right)  - 
     13 \dtw \dth - \frac{25 {\dth}^2}{2} 
\right ]  g^4 \ssep 
  + \left [ \frac{-21 {\don}^3}{4} + 
     {\don}^2 \left( \frac{71 \dtw}{8} - 
        16 \dth \right)  \right ]  g^6 - 
  \frac{19 {\don}^4 g^8}{12}
 \right \}\,,
\label{rgeed3}
\eea
\end{subequations}
where $t$ is defined as $t=\ln (\mu_E)$.
Here we have utilized the modified momentum power counting
rule for the anomalous couplings \cite{yandu}, in which the momentum
dimension of $d_i$ is set to be $-2$, so that the momentum dimension
$d_i g^2$ is zero. Therefore, the terms containing  $d_i g^2$ are treated
similarly as those of the constants in the $\beta$ functions.

In order to compare and contrast, we formulate
the results of the direct integrating-out 
method \cite{Dittmaier} with the decoupling limit 
in its RGE form, which read
\begin{subequations}
\bea
{d g^2\over dt} &=&{g^4\over 8 \pi^2} \left \{
- \frac{29}{4}\right \}\,,\label{rgee0g}\\
{d v\over dt}   &=&{ v \over 8 \pi^2} \left \{
\frac{3 \, g^2}{2}  \right \}\,, \label{rgee0v}\\
{d d_1\over dt} &=&{ 1 \over 8 \pi^2} \left \{
- \frac{1}{12}  \right \}\,, \label{rgee0d1}\\
{d d_2 \over dt}&=&{ 1 \over 8 \pi^2} \left \{
 -\frac{1}{12} \right \}\,, \label{rgee0d2}\\
{d d_3\over dt} &=&{ 1 \over 8 \pi^2} \left \{
-\frac{1}{24}  \right \}\,.
\label{rgee0d3}
\eea
\end{subequations}
To retrieve Eqs. (\ref{rgee0g}-\ref{rgee0d3}) from Eqs. (\ref{rgeeg} -\ref{rgeed3}), 
 we consider the case  where    $d_i$'s are assumed to be of 
the order of $1/(4 \pi)^2$. Substituting these $d_i$'s in right hand side of 
Eqs. (\ref{rgeeg}-\ref{rgeed3}), we find that these  terms are smaller than the 
leading constant terms and thus  they  would correspond to the two 
loop effects. So, keeping the relevance of our calculation up to 
one loop order, we  neglect these terms from higher order loops and 
hence reproduce same set of equations as given in (\ref{rgee0g}-\ref{rgee0d3}).
\section{Numerical analysis}
\label{sec:num}
For the numerical analysis,
we mimic the standard model by choosing the
mass of gauge boson $m_W$ to be $80$ GeV. The Higgs
boson is assumed to be heavier than the gauge bosons
$W$. The initial condition for the coupling $g$ and
the vacuum expectation value $v$ is fixed at the
lower boundary point, $\mu_E=m_W$.
The coupling $g(m_W)$ is chosen to satisfy
\bea
\alpha_g={g^2\over 4\pi}={1\over 30}\,,
\eea
which gives $g(m_W)=0.65$
and the vacuum expectation value is then
fixed by $m_W={1\over 2} g v$, which gives $v(m_W)=247$.
The initial conditions for $d_i$'s $\,i=1,\,2,\,3$
are chosen to be fixed at the matching scale, $\mu_E=m_H$,
as given in Eqs. (\ref{tree1}-\ref{tree3}) and Eqs. (\ref{d11loop}-\ref{d31loop}), 
both at the tree and the one-loop level. 

It is important to clarify that
the masses of gauge boson and Higgs boson
in the ${\overline {MS}}$ and on-shell
renormalization schemes are different.
The relation of the masses of the gauge
boson between the ${\overline {MS}}$ and
the on-shell renormalization scheme up to
one-loop level in the full theory is given as
\bea
m_{M,on-shell}^2 &=& m_W^2 \left [ 1 + {1 \over 16 \pi^2 } 
\left ( {5 r -3 \over 16} +
{r ( 5 - 2 r ) \over 8 (r - 1) } ln (r) - {r + 3 \over 4} ln (r_w) \right ) \right ]\,,
\eea
where $r_W=m_W^2/\mu_E^2$.
However the perturbation theory breaks down for very heavy Higgs boson and
large $r$. 
Therefore, for the sake of simplicity, in the numerical
analysis, we will only 
consider the tree level relations between
the masses of these two schemes, and
assume that the gauge boson's mass is
the one which is measured in the experiments.

We consider three cases to show the effects
of matching conditions. The first case is just the
tree-level matching conditions, the second case is the
one-loop matching conditions in the decoupling limit,
and the third one is the exact one-loop matching conditions.

\begin{figure}[t]
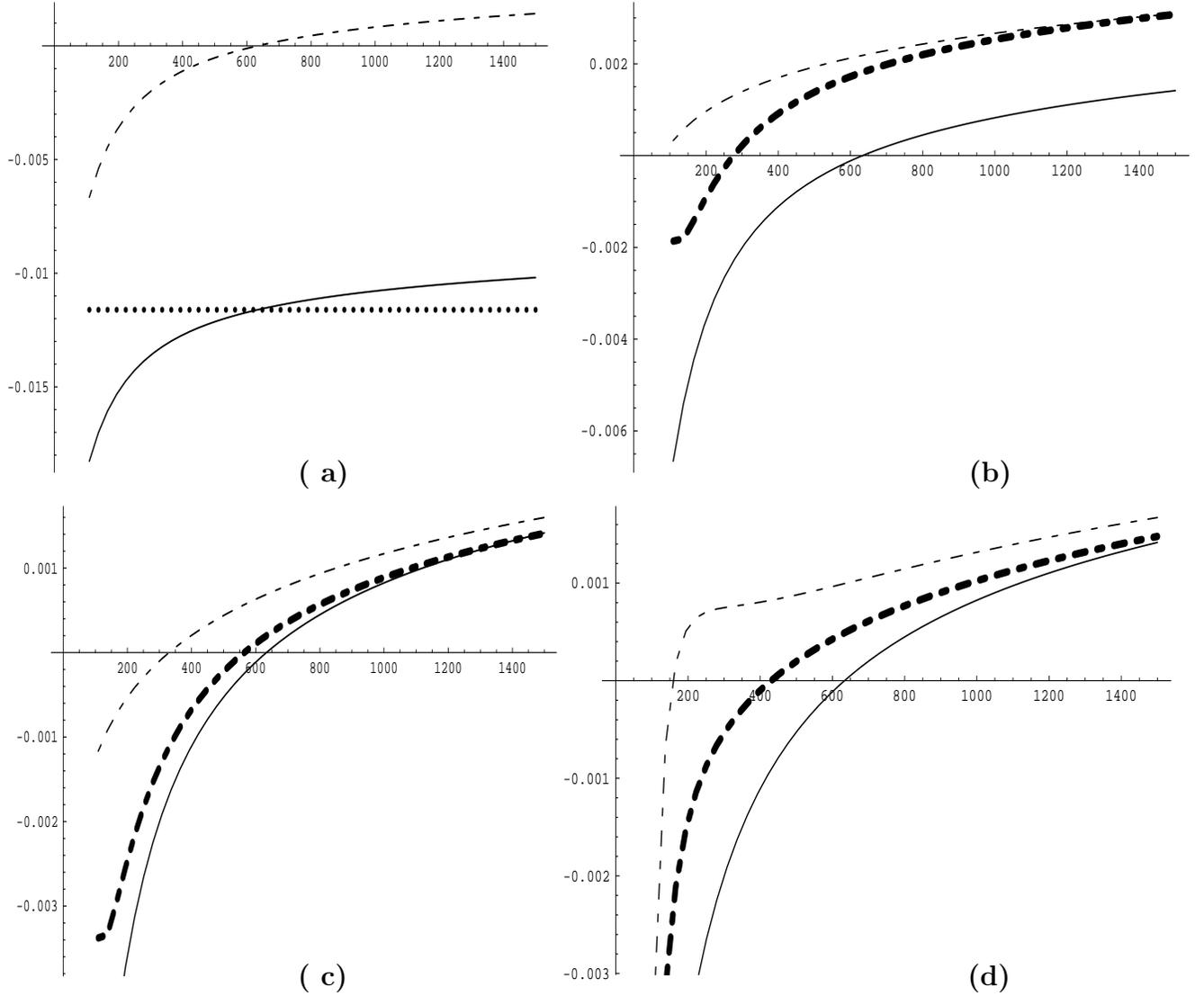

\centerline{
\epsfxsize= 8.0 cm\epsfysize=7.0cm
 \epsfbox{d1cmpw.epsi}
        \hspace*{0.2cm}
\epsfxsize=9.0 cm\epsfysize=7.0cm
                     \epsfbox{d1cmptr.epsi}}
\vskip -.5cm
\hskip 3 cm {\bf ( a) }\hskip 9 cm {\bf (b)}
\vskip 0.2 cm 
\centerline{
\epsfxsize= 8.0 cm\epsfysize=7.0cm
                     \epsfbox{d1cmpdc.epsi}
        \hspace*{-0.2cm}
\epsfxsize=9.0 cm\epsfysize=7.0cm
                     \epsfbox{d1cmplp.epsi}
}
\vskip -.5cm
\hskip 3 cm {\bf ( c) }\hskip 9 cm {\bf(d)}
\vskip 0.5 cm
\caption{\it
 Variation of $d_1 (m_W)$ in $y$ axis is shown with the  $m_H$ (in GeV)
in $x$ axis.
   Figure (a) shows  the constant part (dotted line),  $m_H$
dependent part (dashed line) and the sum of these two (solid line)
from the 1-loop contribution. Figures (b), (c), and (d) corresponds to  tree level,
   1-loop  in the decoupling
limit and  exact  1-loop matching conditions respectively, and  each
of them
depict the comparison between the 
direct integrating-out method (thin dashed line ), the RGE method
 in the effective theory (thick dashed line) and  1-loop 
in the full theory (solid line). 
}
\label{fig2}
\end{figure}

In the Figs. \ref{fig2} , \ref{fig3} and \ref{fig4}
we plot the magnitude of the
effective couplings $d_i$'s with varying  mass of the Higgs 
boson for these three different cases. 

Figs.   \ref{fig2}(a), \ref{fig3}(a) and \ref{fig4}(a) indicate that 
for most range of Higgs mass,
the constant contribution from the gauge and Goldstone
bosons is numerically larger than the contribution from the part
dependent on Higgs boson mass. 

If only considering the non-decoupling log terms alone, Fig. \ref{fig2}(b) shows that 
the direct integrating-out method is
worse than the RGE method when Higgs boson is relatively
light. Fig. \ref{fig2}(c) establishes the fact 
that after taking into account the one-loop  non-decoupling  constants at the matching
scale, both the  methods improve.  
However, the RGE method comes out to be better than the direct integrating-out method.
Fig. \ref{fig2}(d) confirms that after using the exact one-loop matching condition,
the RGE method gives better prediction than the direct integrating-out method for
a wide range of Higgs boson's mass. 

Fig. \ref{fig3}(b) shows that with the tree level matching conditions, 
both the RGE method and the direct integrating 
method deviate from the exact one-loop result significantly.
Fig. \ref{fig3}(c) shows that when taking into account the one-loop
non-decoupling constants, predictions of both methods 
improve. Fig. \ref{fig3}(d) shows when
the exact one-loop matching conditions are used, for a wide range of
Higgs boson the RGE method is better than the
direct integrating-out method. 

Figs. \ref{fig2}(c), \ref{fig2}(d), \ref{fig3}(c) and  Fig. \ref{fig3}(d) 
also show that if Higgs boson's mass becomes heavy, the predictions of 
the direct integrating-out method, the RGE method, and the 
exact one-loop calculation converge. 
While when Higgs boson's mass becomes light, the difference between the effective 
field theory method and the exact one-loop calculation 
become large, which indicates the effects of 
higher order operators in the effective Lagrangian.

\vskip 1cm

\begin{figure}[t]
\centerline{
\epsfxsize= 8.0 cm\epsfysize=7.0cm
 \epsfbox{d2cmpw.epsi}
        \hspace*{0.2cm}
\epsfxsize=9.0 cm\epsfysize=7.0cm
                     \epsfbox{d2cmptr.epsi}}
%\vskip -.5cm
\hskip 3 cm {\bf ( a) }\hskip 9 cm {\bf (b)}
\vskip 0.2 cm 
\centerline{
\epsfxsize= 8.0 cm\epsfysize=7.0cm
                     \epsfbox{d2cmpdc.epsi}
        \hspace*{-0.2cm}
\epsfxsize=9.0 cm\epsfysize=7.0cm
                     \epsfbox{d2cmplp.epsi}
}
%\vskip -.5cm
\hskip 3 cm {\bf ( c) }\hskip 9 cm {\bf(d)}
\vskip 0.5 cm
\caption{\it
 Variation of $d_2 (m_W)$ in $y$ axis is shown with the  $m_H$ (in GeV)
in $x$ axis.
   Figure (a) shows  the constant part (dotted line),  $m_H$
dependent part (dashed line) and the sum of these two (solid line)
from the 1-loop contribution. Figures (b), (c), and (d) corresponds to  tree level,
   1-loop  in the decoupling
limit and  exact  1-loop matching conditions respectively, and  each
of them
depict the comparison between the 
direct integrating-out method (thin dashed line ), the RGE method
 in the effective theory (thick dashed line) and  1-loop 
in the full theory (solid line). 
}
\label{fig3}
\end{figure}
\begin{figure}[t]
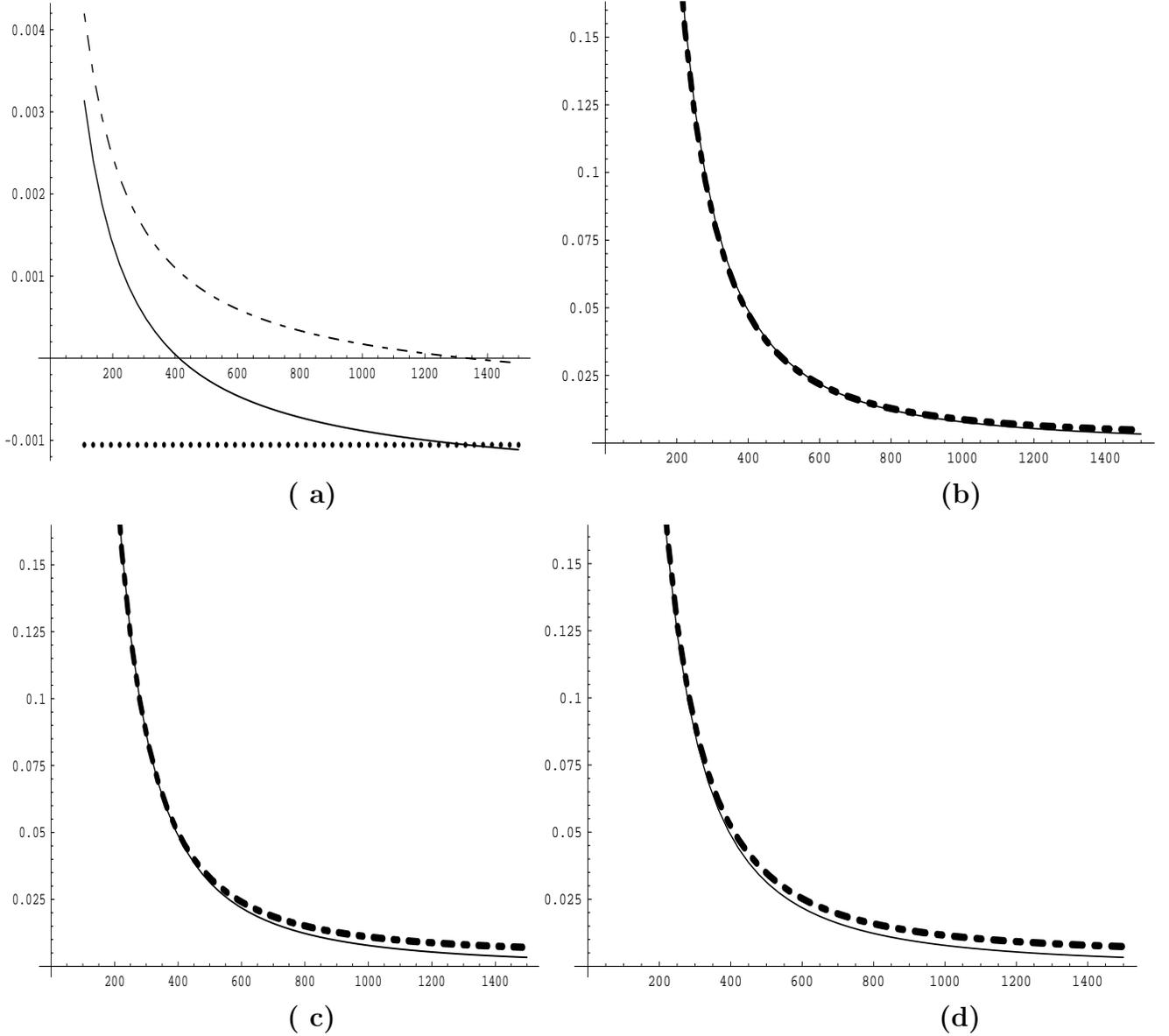

\centerline{
\epsfxsize= 8.0 cm\epsfysize=7.0cm
 \epsfbox{d3cmpw.epsi}
        \hspace*{0.2cm}
\epsfxsize=9.0 cm\epsfysize=7.0cm
                     \epsfbox{d3cmptr.epsi}}
%\vskip -.5cm
\hskip 3 cm {\bf ( a) }\hskip 9 cm {\bf (b)}
\vskip 0.2 cm 
\centerline{
\epsfxsize= 8.0 cm\epsfysize=7.0cm
                     \epsfbox{d3cmpdc.epsi}
        \hspace*{-0.2cm}
\epsfxsize=9.0 cm\epsfysize=7.0cm
                     \epsfbox{d3cmplp.epsi}
}
%\vskip -.5cm
\hskip 3 cm {\bf ( c) }\hskip 9 cm {\bf(d)}
\vskip 0.5 cm
\caption{\it
 Variation of $d_3 (m_W)$ in $y$ axis is shown with the  $m_H$ (in GeV)
in $x$ axis.    Figure (a) shows  the constant part (dotted line),  $m_H$
dependent part (dashed line) and the sum of these two (solid line)
from the 1-loop contribution. Figures (b), (c), and (d) corresponds to  tree level,
   1-loop  in the decoupling
limit and  exact  1-loop matching conditions respectively, and  each
of them
depict the comparison between the 
direct integrating-out method (thin dashed line ), the RGE method
 in the effective theory (thick dashed line) and  1-loop 
in the full theory (solid line). 
}
\label{fig4}
\end{figure}

The tree level contribution of $d_3$ is
overwhelming against the one-loop corrections as shown in Figs.
\ref{fig4}(b), \ref{fig4}(c), and \ref{fig4}(d). When Higgs boson's mass becomes
heavy, the results of the effective field theory method do not converge with
the exact one-loop calculation. The reason for this anomaly might be related with
our approximations we have used.

From these figures, generally speaking, we can draw the 
conclusion that the one-loop matching conditions with 
the one-loop RGE method is reliable to a certain degree, 
and the effective theory description is valid even Higgs boson
is relatively light.
\section{Discussions and Conclusions}
\label{sec:discus}
In this article we study the renormalization of the nonlinear  $SU(2)$
Higgs model in the 
${\overline {MS}}$ renormalization scheme. 
In order to get the divergent structure given
in Eq. (\ref{1ldiv}), we  prescribe an appropriate
parameterization for the Goldstone bosons from the phase angles. We realize
 that the EOMs of both gauge and Higgs bosons are
very crucial in order to verify that the $SU(2)$ Higgs model in its
nonlinear form is renormalizable.

We provide the exact one-loop  calculation of the anomalous couplings
and hence give the matching conditions up to one-loop level. 
It is worth mentioning that these matching conditions are also 
valid for the relatively light Higgs sector.

Our analysis shows that at the one-loop order, the $m_H$ independent part
which comes from the contribution
of gauge and Goldstone bosons is equally important, and bears 
the same order of the magnitude to the part that is dependent on the Higgs boson's mass.
We observe that even for the region where the Higgs boson  is 
relatively light, the effective description is still reliable to
a certain degree. 

Our numerical analysis shows that the predictions of the matching
conditions at the one-loop
level are better than those achieved at the tree-level. By comparing the
predictions of the one-loop RGE and one-loop exact computation,
we find that even for a relatively light Higgs boson, operators up to
$O(p^4)$ in the  gauge boson sector can accounts for  the features of the
Higgs boson in terms of the effective anomalous couplings to a certain
degree.

To compare our results for the anomalous couplings, we have to seek a
proper correspondence between the coordinate space and momentum space 
calculations. Also we have to remember that the existing calculations
in the literature \cite{Dittmaier} accounts only for the $m_H$ dependent  
parts for the very heavy Higgs boson  region. Our results agree with
those given in \cite{Dittmaier} in the decoupling limit if the proper
correspondence is made between the ${\overline {MS}}$  renormalization
scheme in the coordinate space
and the On-Shell renormalization scheme in the momentum space formalisms. 

Before closing, we would like to address  few subtleties and
limitations of  our calculation 
\begin{enumerate}
 \item {\it Gauge operator of the Goldstone bosons}:  The gauge potential
for the Goldstone boson  can be defined as
\bea
{\Gamma_{\mu}^{\xi}} &=& - i {\overline W^{ab}_{\mu}} 
+ \de^{ab} {\partial_{\mu} {\overline H} \over (v + {\overline H}) }\,.
\eea
It is interesting to note that the symmetric part does not contribute to the
antisymmetric field strength tensor. However, this fact renders that 
the trace of the gauge potential defined in the 
gauge covariant differential operator is non-vanishing. 

\item {\it Usage of the EOM} :  In order to eliminate the
linear terms of quantum fields in the Lagrangian,
the EOM of background fields were used. 
We mentioned earlier, the 
feature of such application of
 the EOM (\ref{heom}) to  evaluate  the  finite contributions of the 
term $\sigma_\xi^{ab} $ in the definition of the Goldstone boson's
propagator in Eq. (\ref{sigmaG}).  However, we  have restricted ourselves  from using 
the EOM of the Higgs boson elsewhere. 
For example, the EOM of the Higgs boson is not used in the 
 term  $\rx \Box_{W}^{-1} \lx \Box_{\xi}^{-1}$ in the
effective irreducible vertex functional
in Eq. (\ref{1llag}),  which contains the term $\partial_{\alpha} {\overline H}$
$\partial^{\alpha} \partial^2 {\overline H}$.  This would imply   the
dropping of the higher order terms of the momentum expansion of the
scalar loop integral $B_0(p^2,m^2_1,m^2_2)$. And this is consistent within the
framework of coordinate space calculations.
The use of the EOM of Higgs
does not change the physical amplitudes of the theory as has been
demonstrated in reference \cite{eom1,eom2}.

\item {\it Higher order contributions }: In our computation procedure,
we find that higher order operator affects the calculation 
of the finite contribution of the lower order operator 
$tr[V_{\mu} V^{\mu}] tr[V_{\nu} V^{\nu}]$.
For instance, one of the operator of the order of  $O(p^5)$ 
$\partial_{\alpha} \partial_{\beta} {\overline H} \, 
{\overline W}^{\alpha} {\overline W}^{\beta}$  can be expressed as
\bea
\int \partial_{\alpha} \partial_{\beta} {\overline H} \, {\overline W^{\alpha}} 
{\overline W^{\beta}} &=&
\int {1\over 2} \partial^2 {\overline H} {\overline W} \cdot {\overline W}
- \int {g^2 \over 4} {\overline H} {\overline W^{\beta}} (\partial^2 g_{\alpha \beta} -
\partial_{\beta} \partial_{\alpha}) {\overline W^{\alpha}} \nnb \\ 
&&+ \int {\overline H} {\overline W}_{\mu \nu}^a {\overline W}^{\mu\nu,a}
+ \cdots \,.
\eea
and by using the EOMs of Higgs and gauge  bosons in the above equation,
the first term contribute to $tr[V_{\mu} V^{\mu}] tr[V_{\nu} V^{\nu}]$.
From this example, we realize that the information on the form of the complete set 
of higher order operators 
(say $O(p^5)$, $O(p^6)$, and so on) is necessary in order to
determine the contributions to the lower order operators, and this fact indeed
complicates our computation procedure. This kind of conributions will
affect the value of $d_3$. However, due to this
complication, we omit such contributions from higher order operators.

\item{\it Matching in the coordinate space}:
The matching conditions are constructed to establish the connection
between the parameters of the underlying full theory and the effective
theory.  The consequence of the matching of these two theories directly, 
would  render the couplings of the effective theory non-local. 
In order to extract a local interaction theory, it is customary to expand the
non-local effective couplings as so to get a local interaction 
theory truncated to a specified order. This is,
a common practice for the three and four points functions.
For two point functions, the correspondence is not simple.
In order to realize the On-Shell
renormalization scheme in the coordinate space, we have to sum over
all the higher order terms which contribute to the two point functions, which then 
proves to be a disaster for our computation. However, for the two point
functions, by using the correspondence between the coordinate space
and the momentum space as given in appendix \ref{app:B0},
 we can successfully collect all the relevant terms. 
This makes the On-Shell renormalization scheme workable in
 the coordinate space.
\end{enumerate}

Next we would like to address the more realistic  
$SU(2) \times U(1)$  model in our  project 
of trying to understand the full potential of 
the effective electro-weak Lagrangian.

\section{Acknowledgments}
S. Dutta would  like to thank the 
Science and Engineering Research Council, Department of Science 
and Technology, Government of India for the 
partial financial support.
The work of K. Hagiwara and Q.S. Yan are supported
in part by Grant-in-Aid Scientific Research from 
Ministry of Education, Culture, Science and Technology of Japan.
The work of Q.S. Yan 
is partially supported by the Japan Society for the Promotion of Science
 fellowship program.

\appendix
\def\theequation{\thesection.\arabic{equation}}
\setcounter{equation}{0}

\section{Short distance expansion approximation}
\label{app:heatkernel}
In heat kernel method, the propagators for vector,
Goldstone and Higgs bosons  are defined as 
\begin{subequations}
\bea
\langle x|\Box^{-1,ab}_{W;\mu\nu}|y\rangle&=& \int_0^{\infty} \frac{d
  \tau}{(4 \pi \tau)^{d\over 2}} \exp\left(- m_W^2 \tau \right)
\exp\left( - {z^2\over 4 \tau}\right)
H^{\mu\nu,ab}_{W}(x,y;\tau)\,\,,\label{vpro} \\
\langle x|\Box^{-1,ab}_{\xi}|y \rangle &=& \int_0^{\infty} \frac{d
  \tau}{(4 \pi \tau)^{d \over 2}} \exp\left(- m_W^2 \tau \right)
\exp\left( - {z^2\over 4 \tau}\right) H_{\xi}^{ab}(x,y;\tau)\,\,, \label{spro}\\
\langle x|\Box^{-1}_{H}|y \rangle &=& \int_0^{\infty} \frac{d \tau}{(4 \pi \tau)^{d \over 2}} \exp\left(- m_H^2 \tau \right) \exp\left( - {z^2\over 4 \tau}\right) H_{H}(x,y;\tau)\,\,,
\label{hpro}
\eea
\end{subequations}
where $z=y-x$ is short distance variable. 
The integral over the proper time $\tau$ and the factor
${\exp\left[ - {z^2/(4 \tau)}\right]/(4 \pi \tau)^{d \over 2}}$
contains the quadratic divergent part of the propagator.
The analytic behaviour  of the function  $H_i(x,y;\tau)$ ( where
$ i\,\, \equiv W,\, \xi , h$ )  with reference to the variables
$z$ and $\tau$, enables  its expansion in terms of these variables.
\par The expansion of  $H_i(x,y;\tau)$ in $\tau$ is given as
\bea
H_i(x,y;\tau)&=&H_{i,0}(x,y) +H_{i,1}(x,y) \tau + H_{i,2}(x,y) \tau^2 + \cdots\,,
\eea
where $H_{i,0}(x,y)$, $H_{i,1}(x,y)$, and, $H_{i,2}(x,y)$ are known to be 
Seeley-De Witt coefficients \cite{armadi}. 
The coefficient
$H_0(x,y)$ is the pure
Wilson phase factor, which indicates the
phase change of a quantum state
when propagating from the point $y$ to the point $x$ and
reads
\bea
H_{i,0}(x,y)= C_i \exp \left(- \int_y^x \Gamma_i(z)\cdot dz\right ),
\eea
where $\Gamma_i (z)$ is the affine connection ( dependent on
the group representation of the quantum states ) defined on the
coordinate point $z$ and   $C_i$ is the Lorentz structure related with
the spin state of the \lq$i$\rq \, boson. For gauge bosons  $C_i\equiv g^{\mu\nu}$,
which in Euclidean space is essentially $\delta^{\mu\nu}$, and for
Goldstone and Higgs bosons $C_i=1$. In the coincidence
limit, we have $H_{i,0}(x,y)|_{y\rightarrow x}=1$.

The higher order Seeley-De Witt coefficients are
determined by the recursion relation
\bea
( 1+ n + z_{\mu} D^{\mu}_{i,x} ) H_{i,n+1}(x,y) + (- D_{i,x}^2 + \sigma_i) H_{i,n}(x,y) =0\,.
\label{recr}
\eea
\par  
In the coincidence limit, the second and third Seeley-De Witt coefficients
determined by the recursion relation given in Eq. (\ref{recr}) are expressed as
\begin{subequations}
\bea
H_{i,1}(x) &=& - \sigma_i\,,\label{SDW1}\\
H_{i,2}(x) &=& {1 \over 12} \Gamma_{i,\mu\nu} \Gamma^{\mu\nu}_i\,.\label{SDW2}
\eea
\end{subequations}

\par The short distance expansion of $H_i(x,y;\tau)$
around the coordinate
$x$ is the ordinary Taylor expansion, which
can be expressed as
\bea
H_i(x,y) &=& H_i(x,y)|_{x=y} + z^{\alpha} \partial_{\alpha} H_i(x,y)|_{x=y}
+ {1\over 2} z^{\alpha} z^{\beta} \partial_{\alpha}\partial_{\beta} H_i(x,y)|_{x=y}
+ \cdots\,.
\eea

We make use of this expansion in our
calculation. 

\par We also define the
operation of the covariant differential operator $D_{i,x}^\alpha$ on 
its corresponding propagator defined by $-D_i^2 + m_i^2 + \sigma_i $
in terms of the integral
\bea
\langle x| D_{i,x}^{\alpha} \Box^{-1}_i|y\rangle
&=&\int_{x^{\prime}} \langle x| D_{i,x}^{\alpha} | x^{\prime} \rangle
\langle x^{\prime} |\Box^{-1}_{i}|y\rangle\,\nnb\\
&=&
\int d\lambda {1\over (4 \pi \lambda)^{d/2}} \exp\{-m_i^2 \lambda \} \exp{(- {z^2 \over 4 \lambda})}\nnb\\
&&\times ({z^{\alpha} \over 2 \lambda} + D_i^{\alpha}) H_i(x,y;\lambda)\,.
\eea
Using this definition we can show
\begin{subequations} 
\bea
D^{\alpha}_{i,x} H_{i,0}(x,y)|_{y\rightarrow x}&=&0\,,\label{CSDprop1}\\
D^{\alpha}_{i,x} D^{\beta}_{i,x} H_{i,0}(x,y)|_{y\rightarrow x} &=& {1 \over 2} \Gamma_{i,\alpha \beta}\,.
\label{CSDprop2}
\eea
\end{subequations}

The short distance covariant expansion  of operators  $X(x)$ given in Eqs. (\ref{prop5}-\ref{prop11}) 
are defined as operator $I(x,y)$ which can be expressed as
\bea
I(x,y) &=& H_{i,0}(x,y) X(y) H_{j,0}(y,x)\,\nnb\\
&=& X(x) + z^{\alpha} D_{\alpha} X(x) 
+ {z^{\alpha} z^{\beta} \over 2} D_{\alpha} D_{\beta} X(x) + \cdots \,.
\label{shortex}
\eea
Here $H_{i(j),0}$ means the first Seeley-De Witt coefficient
of the particle $i(j)$, and
\begin{subequations}
\bea
D_{\alpha} X &=& \partial_{\alpha} X + \Gamma_{i, \alpha} X - X \Gamma_{j, \alpha} \,,\label{eqs21}\\
D_{\alpha} D_{\beta} X &=& \partial_{\alpha} \partial_{\beta} X
+  \Gamma_{i, \alpha} \Gamma_{i, \beta} X
+  X \Gamma_{j, \alpha} \Gamma_{j, \beta}
-2 \Gamma_{i, \alpha} X \Gamma_{j, \beta}
\nnb\\&&+2 \Gamma_{i, \alpha} \partial_{\beta} X
-2 \partial_{\alpha} X \Gamma_{j,\beta}
+  \partial_{\alpha} \Gamma_{i,\beta} X
-  X \partial_{\alpha} \Gamma_{j,\beta}\,.
\label{eqs22}
\eea
\end{subequations}
\section {Divergent and finite terms of the Nonlinear SU(2) Higgs Model}
\label{app:div}
In order to extract all the relevant terms up to
$O(p^4)$ (both divergent and finite), we need to
introduce an auxiliary dimension counting rule in coordinate space
calculation, which reads as
\bea
[{\overline W_{\mu}^s}]_a=[\partial_{\mu}]_a=[D_{\mu}]_a=[{\overline H}]_a=1\,,
[v]_a=0.
\label{aucnt}
\eea
The divergence counting rule of the integral
over the coordinate space $z$ and the proper
time $\tau$ can be established as
\bea
[z^{\mu}]_d=1\,\,,\,\,[\tau]_d=2\,.
\label{dcnt}
\eea

The  relevant terms from the trace of single propagator $\ln\Box_i $
given in Eq. (\ref{1llag}) ) are  calculated by standard procedure
and given as
\begin{subequations}
\bea
\ln \Box_{W}&=&- {(\mu^2)^{\left(\epsilon / 2\right)} \over 16 \pi^2} \int_x \left\{
  d \,C_{ad}\, (m_W^2)^{d\over 2} \Gamma\left( - {d\over 2}\right )
-   (m_W^2)^{{d\over 2}-1 } \Gamma\left( {1- {d\over 2}}\right ) \sigma_{W}^{\mu\mu,aa}
\ssep + (m_W^2)^{{d\over 2}-2 } \Gamma\left( {2- {d\over 2}}\right ) \left [
{d C_{ad} \over 12} \Gamma_{\mn} \Gamma^{\mn} + {1\over 2} \sigma_{W}^{\mu\nu,ab}  \sigma_{W}^{\nu\mu,ba}
\right ]
+\cdots
\right \}\,, \label{trlogW}\\
\ln \Box_{\xi}&=&- {(\mu^2)^{\left(\epsilon / 2\right)} \over 16 \pi^2} \int_x \left\{
   C_{ad} \, (m_W^2)^{d\over 2} \Gamma\left( - {d\over 2}\right )
-   (m_W^2)^{{d\over 2}-1 } \Gamma\left( {1- {d\over 2}}\right ) \sigma_{\xi}^{aa}
\ssep + (m_W^2)^{{d\over 2}-2 } \Gamma\left( {2- {d\over 2}}\right ) \left [
{C_{ad} \over 12} \Gamma_{\xi,\mn} \Gamma^{\mn}_{\xi} + {1\over 2} \sigma_{\xi}^{ab}  \sigma_{\xi}^{ba}
\right ]
+\cdots
\right \}\,,\label{trlogG}\\
\ln \Box_{H}&=&- {(\mu^2)^{\left(\epsilon / 2\right)} \over 16 \pi^2} \int_x \left\{
   (m_H^2)^{d\over 2} \Gamma\left( - {d\over 2}\right )
-  (m_H^2)^{{d\over 2}-1 } \Gamma\left( {1- {d\over 2}}\right ) \sigma_{H}
\ssep + (m_H^2)^{{d\over 2}-2 } \Gamma\left( {2- {d\over 2}}\right ) \left [
 {1\over 2} \sigma_{H}^2
\right ]
+\cdots
\right \}\,,\label{trlogH}\\
\ln \Box_{c} &=&- {(\mu^2)^{\left(\epsilon / 2\right)} \over 16 \pi^2} \int_x \left\{
 C_{ad} \, (m_W^2)^{d\over 2} \Gamma\left( - {d\over 2}\right )
- (m_W^2)^{{d\over 2}-1 } \Gamma\left( {1- {d\over 2}}\right ) \sigma_{cc}^{aa}
\ssep + (m_W^2)^{{d\over 2}-2 } \Gamma\left( {2- {d\over 2}}\right ) \left [
{C_{ad} \over 12} \Gamma_{\mn} \Gamma^{\mn} + {1\over 2} \sigma_{cc}^{ab} \sigma_{cc}^{ba}
\right ]
+\cdots
\right \}\,,
\label{trlogGH}
\eea
\end{subequations}
where $C_{ad}=\sum_i {\rm tr}\bigg(  \, t_{ad}^i\,t_{ad}^i\,\bigg)$, and
$t^i_{ad}$ are the matrices of the Lie algebra 
in the adjoint representation.
\par Calculations of the traces with more than one propagator are tedious but
straightforward. However, here we work out the calculation of one trace
 involving  two propagators. For example, consider the term 
$Tr(\rx \Box_{W}^{-1} \lx \Box_{H}^{-1})$.
\bea
S_{WH}&=&Tr(\rx_H \Box_{W}^{-1} \lx_H \Box_{H}^{-1})\nnb\\
&=&\int_x tr (\langle x|\rx_H \Box_{W}^{-1} \lx_H \Box_{H}^{-1} | x\rangle) \nnb \\
&=&\int_x \int_y tr( \rx_H(x) \langle x|\Box_{W}^{-1}| y \rangle \lx_H(y) \langle y|\Box_{H}^{-1}| x \rangle) \nnb\\
&=&\int_x \int_z \int_{\lambda_1} \int_{\lambda_2} 
\exp\left \{ -m_W^2 \lambda_1 - m_H^2 \lambda_2 \right \}
\exp\left \{ -z^2 \left ( {\lambda_1 + \lambda_2 \over 4 \lambda_1 \lambda_2} \right )\right\}\nnb\\
&&\times \rx_H(x)^{\mu,a} H_W^{\mu\nu,ab}(x,y,\lambda_1) \lx_H(y)^{\nu,a} H_{H}(y,x;\lambda_2)\,.
\eea
Using the short distance expansion given in Eq. (\ref{shortex})
and the auxiliary dimension counting rule given in Eq. (\ref{aucnt}),
we can easily find that the relevant terms in $S_{WH}$ can be
expressed as
\bea
S_{WH}&=& \int_x \int_z \int_{\lambda_1} \int_{\lambda_2} 
\exp\left \{ -m_W^2 \lambda_1 - m_H^2 \lambda_2 \right \}
\exp\left \{ -z^2 \left ( {\lambda_1 + \lambda_2 \over 4 \lambda_1 \lambda_2} \right )\right\}
\jsep \times \left \{ \rx_H(x)^{\mu,a} \lx_H(x)^{\mu,a}
+ \lambda_1 \rx_H(x)^{\mu,a} H_{W,1}^{\mu\nu,ab}(x) \lx_H(x)^{\nu,b}
\ssep+ \lambda_2 \rx_H(x)^{\mu,a} \lx_H(x)^{\mu,a} H_{H,1}(x) 
+ {z^{\alpha}z^{\beta} \over 2} \rx_H(x)^{\mu,a} D_{\alpha}^{ab} D_{\beta}^{bc} \lx_H(x)^{\nu,c}
+ (O(p^5))
\right  \}\, .
\eea
After performing the integral over $z$ and proper times $\lambda_1$ and $\lambda_2$,
we arrive at
\bea
S_{W H}&=& l_f \int_x 
\left  \{ 
  \rx_H(x)^{\mu,a} \lx_H(x)^{\mu,a} B_{0,WH}
+ \rx_H(x)^{\mu,a} H_{W,1}^{\mu\nu,ab}(x) \lx_H(x)^{\nu,b} C_{1,W H}
\ssep + \rx_H(x)^{\mu,a} \lx_H(x)^{\mu,a} H_{H,1}^{ca}(x) C_{2,W H}
+ \rx_H(x)^{\mu,a} D_{\alpha}^{ab} D_{\alpha}^{bc} \lx_H(x)^{\nu,c} C_{3,W H}
+ \cdots
\right \}\,,
\eea
where $l_f$ is the one-loop factor $1/(1 6 \pi^2)$.

For all other traces given in Eq. (\ref{1llag})  we list the 
terms contributing up to the order of $O(p^4)$ 
\begin{subequations}
\bea
S_{W\xi}&=&Tr(\rx_{\xi} \Box_{W}^{-1} \lx_{\xi} \Box_{\xi}^{-1})\nnb\\
&=&l_f \int_x 
\left \{ 
  \rx_{\xi}(x)^{\mu,ab} \lx_{\xi}(x)^{\mu,ba} B_{0,W\xi}
+ \rx_{\xi}(x)^{\mu,ab} H_{W,1}^{\mu\nu,bc}(x) \lx_{\xi}(x)^{\nu,ca} C_{1,W \xi}
\ssep + \rx_{\xi}(x)^{\mu,ab} \lx_{\xi}(x)^{\mu,bc} H_{\xi,1}^{ca}(x) C_{2,W \xi}
\ssep + \rx_{\xi}(x)^{\mu,ab} D_{\alpha}^{bc} D_{\alpha}^{cd} \lx_{\xi}(x)^{nu,da} C_{3,W \xi}
+ \cdots
\right \} \,,\label{exptr1}\\
S_{\xi H}&=&Tr(\rx_{\xi} \Box_{\xi}^{-1} \lx_{\xi} \Box_{H}^{-1})\nnb\\
&=& l_f \int_x 
\left  \{ 
  {1 \over 2} \rx(x)^{\alpha,a} \lx(x)^{\alpha,a} A_{0,\xi H}
+ {1 \over 2} \rx(x)^{\alpha,a} H_{\xi,1}^{ab}(x) \lx(x)^{\alpha,b} B_{0,\xi H}
\ssep + {1 \over 2} \rx(x)^{\alpha,a} \lx(x)^{\alpha,a} H_{H,1}(x) B_{1,\xi H}
+{g^{\alpha\beta\alpha'\beta'} \over 2} \rx(x)^{\alpha,a} D_{\alpha^{\prime}}^{ab} D_{\beta^{\prime}}^{bc} \lx(x)^{\beta,c} B_{2,\xi H}
\ssep+ {1 \over 2} \rx(x)^{\alpha,a} \Gamma_{\xi, \alpha \beta}^{ab} \lx(x)^{\beta,b} B_{3, \xi H}
+ \cdots
\right \}\,,\label{exptr2}\\
S_{W \xi H}&=&Tr(\rx_H \Box_{W}^{-1} \lx_{\xi} \Box_{\xi}^{-1} \lx  \Box_{H}^{-1})\nnb\\
&=& l_f \int_x \left  \{ 
  \rx_H(x)^{\mu,a} D_{\alpha}^{ab} \lx_{xi}(x)^{\mu,bc} \rx(x)^{\alpha,c} C_{0,W \xi H}
\ssep + \rx_H(x)^{\mu,a} \lx_{\xi}^{\mu,ab}(x) D_{\alpha}^{bc} \lx(x)^{\alpha,c} C_{1,W \xi H}
\ssep + \rx_H(x)^{\mu,a} \lx_{\xi}(x)^{\mu,a} \rx_{0} C_{2,W \xi H}
+ \cdots
\right \}\,,\label{exptr3}\\
S_{\xi W H}&=&Tr(\rx \Box_{\xi}^{-1} \lx_{\xi} \Box_{W}^{-1} \lx_{H}  \Box_{H}^{-1})\nnb\\
&=& l_f \int_x 
\left  \{ 
  \rx(x)^{\alpha,a} D_{\alpha}^{ab} \lx_{\xi}(x)^{\mu,bc} \lx_H(x)^{\mu,c} C_{0,\xi W H}
\ssep + \rx(x)^{\alpha,a} \lx_{\xi}(x)^{\mu,ab} D_{\alpha}^{bc} \lx_H(x)^{\alpha,c} C_{1,\xi W H}
\ssep + \rx(x)_0 \lx_{\xi}(x)^{\mu,a} \lx_H(x)^{\mu,a} C_{2,\xi W H}
+ \cdots
\right \}\,, \label{exptr4}\\
S_{W\xi W\xi}&=&Tr(\rx_{\xi} \Box_{W}^{-1} \lx_{\xi} \Box_{\xi}^{-1} \rx_{\xi}  \Box_{W}^{-1} \lx_{\xi} \Box_{\xi}^{-1})\nnb\\
&=& l_f \int_x \left \{ 
\rx_{\xi}(x)^{\mu,ab} \lx_{\xi}(x)^{\mu,bc} \rx_{\xi}(x)^{\nu,cd} \lx_{\xi}(x)^{\nu,da} D_{0,W\xi W \xi}
+\cdots
\right \}\,,\label{exptr5}\\
S_{W H W H}&=&Tr(\rx_{H} \Box_{W}^{-1} \lx_{H} \Box_{H}^{-1} \rx_{H}  \Box_{W}^{-1} \lx_{H} \Box_{H}^{-1})\nnb\\
&=& l_f \int_x \left \{ 
\rx_{H}(x)^{\mu,a} \lx_{H}(x)^{\mu,a} \rx_{H}(x)^{\nu,b} \lx_{H}(x)^{\nu,b} D_{0,W H W H}
+\cdots
\right \}\,,\label{exptr6}\\
S_{\xi H \xi H}&=&Tr(\rx \Box_{\xi}^{-1} \lx \Box_{H}^{-1} \rx  \Box_{\xi}^{-1} \lx \Box_{H}^{-1})\nnb\\
&=& l_f \int_x \left \{ 
{g^{\alpha\beta\alpha'\beta'} \over 4} \lx(x)^{\alpha,a} \rx(x)^{\beta,a} \lx(x)^{\alpha',b} \rx(x)^{\beta',b} B_{0,\xi H \xi H}
+\cdots
\right \}\,,\label{exptr7}\\
S_{W \xi W H}&=&Tr(\rx_{H} \Box_{W}^{-1} \lx_{\xi} \Box_{\xi}^{-1} \rx_{\xi}  \Box_{W}^{-1} \lx_H \Box_{H}^{-1})\nnb\\
&=& l_f \int_x \left \{ 
\rx_{H}(x)^{\mu,ab} \lx_{\xi}(x)^{\mu,bc} \rx_{\xi}(x)^{\nu,cd} \lx_{H}(x)^{\nu,d} D_{0,W \xi W H}
+\cdots
\right \}\,,\label{exptr8}\\
S_{\xi W \xi H}&=&Tr(\rx \Box_{\xi}^{-1} \lx_{\xi} \Box_{W}^{-1} \rx_{\xi}  \Box_{\xi}^{-1} \lx \Box_{H}^{-1})\nnb\\
&=& l_f \int_x \left \{ 
\rx(x)^{\alpha,a} \lx_{\xi}(x)^{\mu,ab} \rx_{\xi}(x)^{\mu,bc} \rx(x)^{\alpha,c} C_{0,\xi W \xi H}
+\cdots
\right \}\,,\label{exptr9}\\
S_{\xi H W H}&=&Tr(\rx \Box_{\xi}^{-1} \lx \Box_{H}^{-1} \rx_H  \Box_{W}^{-1} \lx_H \Box_{H}^{-1})\nnb\\
&=& l_f \int_x \left \{ 
\rx(x)^{\alpha,a} \lx(x)^{\alpha,a} \rx_H(x)^{\mu,b} \lx_H(x)^{\mu,b} C_{0,\xi H W H}
+\cdots
\right \}\,.\label{exptr10}
\eea
\end{subequations}
The scalar integrals in Eqs. (\ref{exptr1}-\ref{exptr10}) are given as
\begin{subequations}
\bea
B_{0,WH}&=&\bare
+ \ln({\mu^2 \over m_W^2}) 
+ 1 - {r \over r-1} ln(r) \,,\\
C_{1,WH}&=&{1 \over m_W^2} \left \{ -{1 \over r - 1} + {r \over (r-1)^2 } ln(r) \right\}\,,\\
C_{2,WH}&=&{1 \over m_W^2} \left \{ {1 \over r -1} - {1 \over (r -1)^2} ln(r) \right\}\,,\\
C_{3,WH}&=&{1 \over m_W^2} \left \{ {r+1 \over 2 (r-1)^2} - {r \over (r-1)^3} ln(r)\right\}\,,\\
B_{0,W\xi}&=&\bare
+ \ln({\mu^2 \over m_W^2}) \,,\\
C_{1,W\xi}&=&{1 \over 2 m_W^2}\,,\\
C_{2,W\xi}&=&{1 \over 2 m_W^2}\,,\\
C_{3,W \xi}&=&{1 \over 6 m_W^2}\,,\\
A_{0,\xi H}&=&m_W^2 \left \{
 {(1 + r) \over 2 } \left (\bare
+ \ln({\mu^2 \over m_W^2}) \right )
+ {3 (r +1) \over 4} - {r^2 \over 2 (r-1)} \ln(r) \right \} \,,\\
B_{0,\xi H}&=&- {1 \over 2} \bare 
- {1 \over 2} \ln({\mu^2 \over m_W^2})
+ {1 -3 r \over 4( r-1)} + {r^2 \over 2 (r-1)^2 } \ln(r)\,,\\
B_{1,\xi H}&=&-{1\over 2} \bare 
- {1 \over 2} \ln({\mu^2 \over m_W^2})
+ {3 -r \over 4 ( r-1)} + { r (r-2) \over 2 (r-1)^2} \ln(r)\,,\\
B_{2,\xi H}&=&-{1\over 6} \bare 
- {1 \over 6}  \ln({\mu^2 \over m_W^2})
+ {-5 r^2 + 22 r -5 \over 36 (r -1)^2} 
+ {(r -3 ) r^2 \over 6 ( r-1)^3 } \ln(r)\,,\\
B_{3,\xi H}&=&B_{0,\xi H}\,,\\
C_{0,W \xi  H}&=&{1 \over m_W^2} \left ( 
  {3 r -1 \over 4 (r-1)^2} 
- {r^2 \over 2 ( r-1)^3} \ln(r) \right )\,,\\
C_{1,W \xi  H}&=&{1 \over m_W^2} \left (
{1 -3 r \over 4 (r-1)^2} 
+ {r^2 \over 2 ( r-1)^3} \ln(r) \right )\,,\\
C_{2,W \xi  H}&=&{1 \over m_W^2} \left (
- {1 \over r-1} 
+ {r \over ( r-1)^2} \ln(r) \right )\,,\\
C_{0,\xi W H}&=&{1 \over m_W^2} \left (
- {r-3 \over 4 (r-1)^2} 
+ {r ( r-2) \over 2 ( r-1)^3} \ln(r) \right )\,,\\
C_{1,\xi W H}&=&{1 \over m_W^2} \left (
 {3r -1 \over 4 (r-1)^2} 
- {r^2 \over 2 ( r-1)^3} \ln(r) \right )\,,\\
C_{2,\xi W H}&=&C_{2, W \xi H} \,,\\
D_{0,W\xi W \xi}&=&{1 \over m_W^4} {1 \over 6} \,,\\
D_{0,W H W H}&=&{1 \over m_W^4} \left (
- {2 \over (r-1)^2} 
+ { r+1 \over (r-1)^3} \ln (r) \right)\,,\\
B_{0,\xi H \xi H}&=&{1\over 6} \bare
+ {1\over 6} \ln \left ( {\mu^2 \over m_W^2} \right )
-{-5 r^2 + 22 r - 5 \over 36 ( r-1)^2} - {r^2 (r-3) \over 6 (r-1)^3} \ln(r)\,,\\
D_{0,W \xi W H}&=&{1 \over m_W^4} \left (
{r+ 1 \over 2 (r-1)^2} - {r \over ( r-1)^3} \ln(r)
\right )
\,,\\
C_{0,\xi W \xi H}&=&{1 \over m_W^2} \left (
{ 1 - 3 r \over 4 (r-1)^2} + {r^2 \over 2 ( r-1)^3} \ln(r)
\right )\,,\\
C_{0,\xi H W H}&=&{1 \over m_W^2} \left (
{ r+1\over 2 (r-1)^2} - {r \over ( r-1)^3} \ln(r)
\right )\,,
\eea
\end{subequations}
here $A,\, B,\, C$ and $D$  are the scalar integrals.

\section{ Divergences of the one-loop  SU(2) chiral Lagrangian}
\label{app:rge}
In reference \cite{yandu} authors have provided the details.
We found couple of errors in this calculation and here we rectify
them. Following the BFM we split   the
vector and Goldstone fields of the chiral 
Lagrangian as given in Eq. (\ref{effl})
into classical and quantum parts.
We collect the mixing operators
between quantum vector bosons and Goldstone bosons into the 
standard form, and then we can get the one-loop irreducible
generating functional which is expressed as

\bea
\Gamma^{chl}_{1-loop} &=& - { 1\over 2 }\,\, \,  \left[
Tr\ln \Box_W + Tr\ln \Box_{\xi} - 2 Tr\ln \Box_c 
\ssep
+ Tr(X^{\alpha\beta} d_{\alpha} d_{\beta} \Box_{\xi}^{-1})
-{1 \over 2} Tr(X^{\alpha\beta} d_{\alpha} d_{\beta} \Box_{\xi}^{-1}
X^{\alpha'\beta'} d_{\alpha'} d_{\beta'} \Box_{\xi}^{-1})
\ssep -Tr(\rx_{\xi} \Box_{W}^{-1} \lx_{\xi} \Box_{\xi}^{-1})
+ Tr(\rx_{\xi} \Box_{W}^{-1} \lx_{\xi} \Box_{\xi}^{-1} X^{\alpha\beta} d_{\alpha} d_{\beta} \Box_{\xi}^{-1} )
\ssep
-{1 \over 2} Tr(\rx_{\xi} \Box_{W}^{-1} \lx_{\xi} 
\Box_{\xi}^{-1} \rx_{\xi} \Box_{W}^{-1} \lx_{\xi} \Box_{\xi}^{-1})
+ \cdots
\right]\,.
\label{su2chl}
\eea
 While 
operators in Eq. (\ref{su2chl}) are defined as
\begin{subequations}
\bea
\Box^{\mu\nu,ab}_{W\, W} &=& \left (- D'^{2,ab} + m_W^2 \delta^{ab}
\right) g^{\mu\nu} + \sigma_{WW}^{\mu\nu,ab}\,\, ,\label{stdfe1}\\
\Box_{\xi\,\xi}'^{ab} &=& \Box_{\xi\,\xi}^{ab} + X^{\alpha,ac}
d_{\alpha}^{cb} + X^{\alpha\beta,ac} d_{\alpha}^{cd}
d_{\beta}^{db}\,\, , \label{stdfe2}\\
\Box_{\xi\,\xi}^{ab}&=& \left(- d^{2,ab} + \delta^{ab} m_W^2\right ) +
\sigma_{2,\xi\xi}^{ab} + \sigma_{4,\xi\xi}^{ab}\, ,\label{stdfe3}\\
\Box_{{\bar c}c}^{ab}&=& \left(- D'^{2,ab} + m_W^2 \delta^{ab}
\right)\, , \label{stdfe4}\\
 {\stackrel{\leftharpoonup}{X}}^{\mu,ab}&=&
 {\stackrel{\leftharpoonup}{X}}^{\mu,ac}_{\alpha\beta} d^{\alpha,cd} d^{\beta,db}
+{\stackrel{\leftharpoonup}{X}}^{\mu\alpha,ac} d_{\alpha}^{cb}
+{\stackrel{\leftharpoonup}{X}}^{\mu,ab}_{01}
+{\stackrel{\leftharpoonup}{X}}^{\mu,ab}_{03Z}
+\partial_{\alpha}
{\stackrel{\leftharpoonup}{X}}^{\mu\alpha,ab}_{03Y}\, , \label{stdfe5}\\
 {\stackrel{\rightharpoonup}{X}}^{\nu,ab}&=&
 {\stackrel{\rightharpoonup}{X}}^{\nu,ac}_{\alpha\beta} D'^{\alpha,cd} D'^{\beta,db}
+{\stackrel{\rightharpoonup}{X}}^{\nu\alpha,ac} D'^{cb}_{\alpha}
+{\stackrel{\rightharpoonup}{X}}^{\nu,ab}_{01}
+{\stackrel{\rightharpoonup}{X}}^{\nu,ab}_{03Z}
+\partial_{\alpha} {\stackrel{\rightharpoonup}{X}}^{\nu\alpha,ab}_{03Y}\,,
\label{stdfe6}
\eea
\end{subequations}
the operators appearing in Eqs. (\ref{stdfe1}-\ref{stdfe6}) are defined in \cite{yandu}.

Since the trace log terms are similar to those given in
Eqs. (\ref{trlogW}-\ref{trlogGH}), below we list the divergent structures 
of the rest of terms in Eq. (\ref{su2chl}),
\begin{subequations}
\bea
S_{\xi,1}&=&Tr(X^{\alpha\beta} d_{\alpha} d_{\beta} \Box_{\xi}^{-1})\nnb\\
&=& - {1 \over 2} l_f \int_x \left \{ 
  g^{\alpha\beta} tr(X^{\alpha\beta}) Q_{4}
+ g^{\alpha\beta} tr(X^{\alpha\beta} H_{\xi,1}) Q_2
\right \} \,, \label{mixt1}\\
S_{\xi,2}&=&Tr(X^{\alpha\beta} d_{\alpha} d_{\beta} \Box_{\xi}^{-1}
X^{\alpha'\beta'} d_{\alpha'} d_{\beta'} \Box_{\xi}^{-1})\nnb\\
&=&l_f \int_x \left \{ 
{g^{\alpha\beta\alpha'\beta'} \over 4} tr(X^{\alpha\beta} X^{\alpha'\beta'})
Q_4
\right\}\,, \label{mixt2}\\
S_{v\xi\xi}&=&Tr(\rx_{\xi} \Box_{W}^{-1} \lx_{\xi} \Box_{\xi}^{-1} X^{\alpha\beta} d_{\alpha} d_{\beta} \Box_{\xi}^{-1} )\nnb\\
&=&l_f \int_x \left\{
-{g^{\alpha\beta\alpha'\beta'\alpha''\beta''} \over 8} tr(\rx^{\mu,\alpha\beta} \lx^{\mu,\alpha'\beta'} X^{\alpha''\beta''}) Q_4
\ssep +{g^{\alpha\beta\alpha'\beta'} \over 4} tr(\rx^{\mu,\alpha\beta} \lx^{\mu}_{01} X^{\alpha'\beta'} 
+ \rx^{\mu}_{01} \lx^{\mu,\alpha\beta} X^{\alpha'\beta'}) Q_2
-{g^{\alpha\beta} \over 2} tr(\rx^{\mu}_{01} \lx^{\mu}_{01} X^{\alpha\beta}) Q_0
\right \} \,,\label{mixt3} \\
S_{v\xi v\xi}&=&Tr(\rx_{\xi} \Box_{W}^{-1} \lx_{\xi} 
\Box_{\xi}^{-1} \rx_{\xi} \Box_{W}^{-1} \lx_{\xi} \Box_{\xi}^{-1})\nnb\\
&=&l_f \int_x \left \{ 
{g^{\alpha\beta\alpha'\beta'\alpha''\beta''\alpha'''\beta'''} \over 16} tr(
\rx^{\mu,\alpha\beta} \lx^{\mu,\alpha'\beta'} \rx^{\nu,\alpha''\beta''} \lx^{\nu,\alpha'''\beta''''}) Q_4
\ssep - {g^{\alpha\beta\alpha'\beta'\alpha''\beta''}\over 8} tr(
  \rx^{\mu,\alpha\beta} \lx^{\mu,\alpha'\beta'} \rx^{\nu,\alpha''\beta''} \lx^{\nu}_{01}
+ \rx^{\mu,\alpha\beta} \lx^{\mu,\alpha'\beta'} \rx^{\nu}_{01} \lx^{\nu,\alpha''\beta''}
\ssep + \rx^{\mu,\alpha\beta} \lx^{\mu}_{01} \rx^{\nu,\alpha'\beta'} \lx^{\nu,\alpha''\beta''}
+ \rx^{\mu}_{01} \lx^{\mu,\alpha\beta} \rx^{\nu,\alpha'\beta'} \lx^{\nu,\alpha''\beta''}
) Q_2
\ssep + {g^{\alpha\beta\alpha'\beta'} \over 4} tr(
  \rx^{\mu,\alpha\beta} \lx^{\mu,\alpha'\beta'} \rx^{\nu}_{01} \lx^{\nu}_{01}
+ \rx^{\mu}_{01} \lx^{\mu}_{01} \rx^{\nu,\alpha\beta} \lx^{\nu,\alpha'\beta'}
\ssep + \rx^{\mu,\alpha\beta} \lx^{\mu}_{01} \rx^{\nu,\alpha'\beta'} \lx^{\nu}_{01}
+ \rx^{\mu}_{01} \lx^{\mu,\alpha\beta} \rx^{\nu}_{01} \lx^{\nu,\alpha'\beta'} 
\ssep + \rx^{\mu,\alpha\beta} \lx^{\mu}_{01} \rx^{\nu}_{01} \lx^{\nu,\alpha'\beta'}
+ \rx^{\mu}_{01} \lx^{\mu,\alpha\beta} \rx^{\nu,\alpha'\beta'} \lx^{\nu}_{01}
) Q_0
\right\}\,,\label{mixt4}\\
S_{v\xi} &=& Tr(\rx_{\xi} \Box_{W}^{-1} \lx_{\xi} \Box_{\xi}^{-1})\nnb\\
&=&l_f \int_x \left \{
t_{AA} + t_{BB} + t_{CC} + 
t_{AB} + t_{AC} + t_{BC}
\right \}\,,\nnb\\
t_{AA}&=&{1 \over 4} g^{\alpha\beta\alpha'\beta'} g^{\mn} 
tr[{\stackrel{\rightharpoonup}{X}}^{\mu}_{\alpha\beta}
   {\stackrel{\leftharpoonup}{X}}^{\nu}_{\alpha'\beta'}] Q_4 \nnb\\
&&+ {g^{\mn}\over 4} ({g^{\alpha\beta\alpha'\beta'\delta\gamma} \over 6} - { g^{\alpha\beta} g^{\alpha'\beta'\delta\gamma} \over 2}
-{g^{\alpha'\beta'} g^{\alpha\beta\delta\gamma} \over 2}+ g^{\alpha\beta} g^{\alpha'\beta'} g^{\delta\gamma})
tr[{\stackrel{\rightharpoonup}{X}}^{\mu}_{\alpha\beta} D_{\delta}D_{\gamma}{\stackrel{\leftharpoonup}{X}}^{\nu}_{\alpha'\beta'}] Q_2\nnb\\
&& + {g^{\alpha\beta\alpha'\beta'} \over 8} \left [
 g_{\mn} tr[{\stackrel{\rightharpoonup}{X}}^{\mu}_{\alpha\beta}{\stackrel{\leftharpoonup}{X}}^{\nu}_{\alpha'\beta'} H_{\xi,1}]
+ g_{\mu\mu'} g_{\nu\nu'} tr[{\stackrel{\rightharpoonup}{X}}^{\mu}_{\alpha\beta}H_{W,1}^{\mu'\nu'}{\stackrel{\leftharpoonup}{X}}^{\nu}_{\alpha'\beta'}]
\right ] Q_2 \,,\nnb\\
t_{AB}&=& { g_{\alpha'\alpha''} g_{\mu\nu} \over 2} (g^{\alpha\beta}g^{\alpha'\beta'} -{1\over 2} g^{\alpha\beta\alpha'\beta'} )
tr[{\stackrel{\rightharpoonup}{X}}^{\mu}_{\alpha\beta} D_{\beta'} {\stackrel{\leftharpoonup}{X}}^{\nu\alpha''}
- {\stackrel{\rightharpoonup}{X}}^{\mu\alpha''} D_{\beta'} {\stackrel{\leftharpoonup}{X}}^{\nu}_{\alpha\beta}] Q_2\,,\nnb\\
t_{AC}&=&  - {g^{\alpha\beta} g_{\mn} \over 2}
tr[{\stackrel{\rightharpoonup}{X}}^{\mu}_{\alpha\beta} {\stackrel{\leftharpoonup}{X}}^{\nu}_{01}
+{\stackrel{\rightharpoonup}{X}}^{\mu}_{01} {\stackrel{\leftharpoonup}{X}}^{\nu}_{\alpha\beta}
+{\stackrel{\rightharpoonup}{X}}^{\mu}_{\alpha\beta} {\stackrel{\leftharpoonup}{X}}^{\nu}_{03Z}
+{\stackrel{\rightharpoonup}{X}}^{\mu}_{03Z} {\stackrel{\leftharpoonup}{X}}^{\nu}_{\alpha\beta}\nnb\\
&&-\partial_{\alpha'} {\stackrel{\rightharpoonup}{X}}^{\mu}_{\alpha\beta} {\stackrel{\leftharpoonup}{X}}^{\nu\alpha'}_{03Y}
-{\stackrel{\rightharpoonup}{X}}^{\mu\alpha'}_{03Y} \partial_{\alpha'} {\stackrel{\leftharpoonup}{X}}^{\nu}_{\alpha\beta}] Q_2 \nnb\\
&&-{1\over 4} g^{\alpha\beta}
 tr[
   {\stackrel{\rightharpoonup}{X}}^{\mu}_{01} {\stackrel{\leftharpoonup}{X}}^{\mu}_{\alpha\beta} H_{\xi,1}
+  {\stackrel{\rightharpoonup}{X}}^{\mu}_{01} H^{\mu\nu}_{W,1} {\stackrel{\leftharpoonup}{X}}^{\nu}_{\alpha\beta}
+  {\stackrel{\rightharpoonup}{X}}^{\mu}_{\alpha\beta} {\stackrel{\leftharpoonup}{X}}^{\mu}_{01}  H_{\xi,1} 
+  {\stackrel{\rightharpoonup}{X}}^{\mu}_{\alpha\beta} H^{\mn\nu}_{W,1} {\stackrel{\leftharpoonup}{X}}^{\nu}_{01}
] Q_0\,\nnb\\
&& + g_{\mu\nu} ({1\over 6} g^{\alpha\beta\alpha'\beta'} - {1\over 4} g^{\alpha\beta} g^{\alpha'\beta'})
tr [{\stackrel{\rightharpoonup}{X}}^{\mu}_{\alpha\beta} D_{\alpha'}D_{\beta'}{\stackrel{\leftharpoonup}{X}}^{\nu}_{01}
+{\stackrel{\rightharpoonup}{X}}^{\mu}_{01} D_{\alpha'}D_{\beta'}{\stackrel{\leftharpoonup}{X}}^{\nu}_{\alpha\beta}] Q_0\,,\nnb\\
t_{BB}&=&- {g_{\mn} g_{\alpha\beta} \over 2} tr[{\stackrel{\rightharpoonup}{X}}^{\mu\alpha} {\stackrel{\leftharpoonup}{X}}^{\nu\beta}] Q_2\,,\nnb\\
t_{BC}&=& { g^{\alpha\beta} g_{\alpha\alpha'} g_{\mn} \over 2}
tr[{\stackrel{\rightharpoonup}{X}}^{\mu\alpha'} D_{\beta} {\stackrel{\leftharpoonup}{X}}^{\nu}_{01}
-{\stackrel{\rightharpoonup}{X}}^{\mu}_{01} D_{\beta} {\stackrel{\leftharpoonup}{X}}^{\nu\alpha'}] Q_0 \,,\nnb\\
t_{CC}&=&g_{\mn} tr[{\stackrel{\rightharpoonup}{X}}^{\mu}_{01} {\stackrel{\leftharpoonup}{X}}^{\nu}_{01}
+{\stackrel{\rightharpoonup}{X}}^{\mu}_{01} {\stackrel{\leftharpoonup}{X}}^{\nu}_{03Z}
+{\stackrel{\rightharpoonup}{X}}^{\mu}_{03Z} {\stackrel{\leftharpoonup}{X}}^{\nu}_{01}
-\partial_{\alpha'} {\stackrel{\rightharpoonup}{X}}^{\mu}_{01} {\stackrel{\leftharpoonup}{X}}^{\nu\alpha'}_{03Y}
-{\stackrel{\rightharpoonup}{X}}^{\mu\alpha'}_{03Y} \partial_{\alpha'} {\stackrel{\leftharpoonup}{X}}^{\nu}_{01}] Q_0\,.
\label{mixt5}
\eea
\end{subequations}
The divergent functions $Q_i$ represent quartic, quadratic
and logarithmic divergences, respectively, which are given as
\begin{subequations}
\bea
Q_4 &=& (m_W^2) ^{{d \over 2} } \Gamma\left( - {d \over 2} \right)\,\nnb\\
&=&{m_W^4 \over 2} \left \{  {2 \over \epsilon} + {3 \over 2}  + \ln({\mu^2 \over m_W^2})
\right \}\,,\label{divquart}\\
Q_2 &=& (m_W^2) ^{{d \over 2} - 1} \Gamma \left(1 - {d \over 2} \right)\nnb\\
&=& - m_W^2 \left \{  {2 \over \epsilon} +1  + \ln\left({\mu^2 \over m_W^2}\right)
\right \} \,,\label{divquad}\\
Q_0 &=& (m_W^2) ^{{d \over 2} - 2} \Gamma \left(2 - {d \over 2} \right)\,\nnb\\
&=& {2 \over \epsilon}  + \ln\left({\mu^2 \over m_W^2}\right)\,.\label{divlog}
\eea
\end{subequations}
The terms  ${\stackrel{\rightharpoonup}{X}}^{\mu}_{01} H^{\mu\nu}_{W,1} {\stackrel{\leftharpoonup}{X}}^{\nu}_{\alpha\beta}$
and ${\stackrel{\rightharpoonup}{X}}^{\mu}_{\alpha\beta}
{\stackrel{\leftharpoonup}{X}}^{\mu}_{01}  H_{\xi,1}$ in  $t_{AC}$
given in Eq. (\ref{mixt5}) are
missing in the reference \cite{yandu}, which will modified the $\beta$ functions of
$d_i$. However, this modification does not change the conclusion of reference \cite{yandu}.

\section{Scalar loop  integrals in the coordinate space}
\label{app:B0}
Here we discuss the interpretation of scalar $B_0$ integrals in context
of  the coordinate space calculations and compare it with the 
momentum space calculations.

In order to establish the correspondence of a scalar integral,
 here we consider the real scalar $\phi^4$ theory.
 The Lagrangian of this model is given as 
\bea
 L&=&{1 \over 2} \partial \phi \cdot \partial \phi 
+ {1\over 2} \phi^2 + {\lambda \over 4 \!} \phi^4\,.
\eea
In the BFM, we decompose 
$\phi$ into classical and quantum parts as
$\phi={\overline \phi} + \widehat \phi$.
Since we only consider the one-loop corrections,
the only relevant terms in the Lagrangian is
the quadratic terms, which can be formulated as
\bea
 L_{quad}&=&{1 \over 2} {\widehat \phi} 
\Box \widehat \phi\,, \nnb\\
\Box&=& - \partial^2 + m^2 + \sigma \,,\nnb\\
\sigma&=& {1 \over 2} {\overline \phi}^2
 \,.
\eea

After performing the path integral, the one-loop effective
Lagrangian can be represented as
\bea
 L^{1-loop}_{eff} &=&- {1 \over 2} Tr \ln \Box\,.
\eea
To evaluate $Tr \ln \Box$, by using the heat kernel method
in coordinate space, we have
\bea
Tr \ln \Box&=& \int_x \ln [- \partial^2 + m^2 + \sigma ]\,\nnb\\
&=& \int_x \ln [- \partial^2 + m^2] 
+ \int_x \ln[1 + \sigma (- \partial^2 + m^2)^{(-1)}]\nnb\\
&=&\int_x \ln [- \partial^2 + m^2] 
- \int_x \sigma (- \partial^2 + m^2)^{(-1)}
\jsep + {1 \over 2} \int_x \sigma (- \partial^2 + m^2)^{(-1) } \sigma (- \partial^2 + m^2)^{(-1) }
+ \cdots \nnb\\
&=&l_f  \left \{  (m^2)^{d\over 2} \Gamma(-{d \over 2})
- \int_x \sigma(x) A(m^2)
+ {1 \over 2} \int_x \sigma^2(x) B_0(0,m^2,m^2)
+\cdots
\right \}\,.
\label{xspace}
\eea

Now we evaluate $Tr \ln \Box$ in momentum space, 
\bea
Tr \ln \Box &=& \int_x \ln [- \partial^2 + m^2] 
- \int_x \sigma \,\,\, (- \partial^2 + m^2)^{(-1)}
\jsep + {1 \over 2} \int_x \sigma (- \partial^2 + m^2)^{(-1) } \sigma (- \partial^2 + m^2)^{(-1) }
+ \cdots \nnb\\
&=&\int_k \ln [k^2 + m^2]
- \int_k {\tilde \sigma} (k^2 + m^2)^{(-1)}
\jsep + {1 \over 2} \int_k {\tilde \sigma}(k_1,p-k_1) (k^2 + m^2)^{(-1)} 
{\tilde \sigma}(k_1', -p -k_1') ((k+p) + m^2)^{(-1)}
+\cdots \nnb\\
&=&l_f \left \{  (m^2)^{d\over 2} \Gamma(-{d \over 2})
- \int_{k_1} {\tilde \sigma}(k_1,- k_1) A(m^2)
\ssep + {1 \over 2} \int_{k_1} \int_{k_1'} \int_{p} {\tilde \sigma}(k_1,p-k_1) {\tilde \sigma}(k_1',-p-k_1') B_0(p^2,m^2,m^2)
+\cdots
\right \}\,,
\label{pspace}
\eea
where ${\tilde \sigma}$ is given
as
\bea
{\tilde \sigma}(k_1,k_2) &=& 
\int_{k_1} \int_{k_2} {\overline \phi}(k_1) {\overline \phi}(k_2)
\delta^{d}(k_1 + k_2 + p_1 + p_2)\, .
\eea
Here $p_1$ and $p_2$  are the momentum of the quantum fields $\widehat\phi$  respectively.
${\overline \phi}(k_1)$ is the Fourier transformation of the background field
${\overline \phi}$.

We find that if we expand the $B_0(p^2,m^2,m^2)$ 
given in Eq.(\ref{pspace}) with reference to $p^2$,
which appeared in the momentum space calculation,
the $B(0,m^2,m^2)$ given in Eq.(\ref{xspace}) is essentially the
first term of this expansion, which appeared 
in the coordinate space calculation.

Here we also provide some results of this scalar integral $B_0(p^2,m_1^2, m_2^2)$ \cite{denner}, which
are used in section \ref{sec:treematch} to make one to one correspondence to the
results of reference \cite{Dittmaier} in the decoupling limit.
 $B_0(p^2,m_1^2, m_2^2)$ can be
exactly expressed as
\bea
B_0(p^2,m_1^2, m_2^2) &=& \bare + 2 - \ln \left({m_1 m_2 \over \mu^2}\right)
+ {m_1^2 -m_2^2 \over p^2} \ln\left({m_1 \over m_2}\right)
-{m_1 m_2 \over p^2} \left ( {1 \over r} - r \right )
\ln (r)\,,
\eea
where $r$ satisfies
\bea
x^2 + {m_1^2 m_2^2 - p^2 - i\epsilon \over m_1 m_2} x + 1 = (x +r ) \left(x + {1 \over r}\right)\,.
\eea

Here we list the values of $B_0(p^2,m_1^2, m_2^2)$ in some special cases that
we have used in evaluating the anomalous couplings at the one loop level:
\begin{enumerate}
\begin{subequations}
\item[] Case 1:  In the coordinate space calculation, as
  we have learnt that a scalar one loop $B_0$ integral can be defined with
  external momentum $p^2=0$ and  internal masses $m_i, {i=1,2}$ as 
\bea
B_0(0,m_1^2, m_2^2)&=& \bare + 1 - {r_2 \ln r_2 - r_1 \ln r_1 \over r_2 -r_1}\,,
\eea
where $r_i=m_i^2/\mu^2_E$. 
\item[] Case 2: In the coordinate space, the contribution from the Figs.
  \ref{fig1}(a),\ref{fig1}(b) and \ref{fig1}(c)  can be realized with
 $p^2=0$, and $m_1=m_2$ as
\bea
B_0(0,m_1^2, m_1^2)&=& \bare - \ln r_1\,.
\eea
\item[] Case 3: However, in the momentum space, working in Landau gauge
  the scalar integral corresponding to the Fig. \ref{fig1}(d) can be realized
with  $m_1=m_2=0$ as
\bea
B_0(p^2,0,0)&=& \bare - \ln \left({p^2 \over \mu^2_E}\right) + 2 + i \pi \,.
\eea
\item[] Case 4: Similarly, in the momentum space, working in On-Shell
  renormalization scheme would render the contribution from the Fig.
  \ref{fig1}(e) with
 $p^2 = m_1^2 = m_2^2 = m^2$ as
\bea
B_0(m^2,m^2,m^2)&=& \bare - \ln \left({m^2 \over \mu^2_E}\right) + 2 - {\pi \over \sqrt{3} } \,.
\eea
\end{subequations}
\end{enumerate}

\end{document}